\documentclass[twocolumn]{aastex631}

\usepackage{amsmath}
\usepackage{enumitem} 

\shorttitle{Featureless Spectra from Luminous Transients}
\shortauthors{}
\graphicspath{{./}{}}

\begin{document}

\title{The Emission and Suppression of Line Features in Luminous Transients}

\correspondingauthor{Olivia Aspegren}
\email{oliviaaspegren@berkeley.edu}

\author[0000-0001-5674-8403]{Olivia Aspegren}
\affiliation{Department of Astronomy, University of California, Berkeley, CA 94720, USA}

\author[0000-0002-5981-1022]{Daniel Kasen}
\affiliation{Department of Astronomy, University of California, Berkeley, CA 94720, USA}
\affiliation{Department of Physics, University of California, Berkeley, CA 94720, USA}
\affiliation{Nuclear Science Division, Lawrence Berkeley National Laboratory, Berkeley, CA 94720, USA}

\begin{abstract}
Featureless optical and ultraviolet (UV) spectra are a puzzling signature to emerge from recent observations of luminous fast blue optical transients (LFBOTs) and some tidal disruption events (TDEs). We describe the landscape of source and gas properties that are expected to form H, \ion{He}{1} and \ion{He}{2} emission lines, and map spectral types to the parameter space of luminosity and system radius.  Using one-dimensional radiative transfer calculations, we show that high source luminosities ($L > 10^{44}\,\rm erg~s^{-1}$) and compact ejecta radii ($r < 10^{14}\,\rm cm$) produce featureless spectra due to the high temperature and ionization state of the emitting medium. Intermediate luminosities and moderately compact systems can generate \ion{He}{2}-dominated spectra, while lower luminosities and more extended atmospheres result in conspicuous H and \ion{He}{1} emission. Large expansion velocities ($v \geq 0.1c$) can further broaden lines such that they blend into the continuum. Featureless UV spectra may require even more extreme ionization environments or velocities to suppress the many intrinsically strong metal lines at those wavelengths. Applying this framework to understand the absence of features observed in LFBOTs and featureless TDEs, we find that under the optically thick, quasi-thermal conditions considered here, non-homologous, compact outflows are likely necessary for featurelessness to persist in optical and UV spectra.
\end{abstract}


\section{Introduction} \label{sec:intro}

Recent observations have revealed a growing number of luminous transients with surprisingly featureless optical and ultraviolet (UV) spectra. Among the objects that sustain a smooth blackbody continuum with weak or absent spectral features are luminous fast blue optical transients \citep[LFBOTs,][]{margutti_cow, ho_2023} and some tidal disruption events \citep[TDEs,][]{hammerstein_2023, 2023ApJ...955L...6Y}. The physical conditions responsible for the suppression of line features in these sources are poorly understood.

The LFBOTs, a subclass of the fast blue optical transients \citep[FBOTs,][]{2014ApJ...794...23D, 2016ApJ...819...35A} with bolometric luminosities exceeding $10^{43}\,\rm erg~s^{-1}$, persist in near featurelessness despite their rapid rise over a few days and subsequent decline in luminosity. To date, three well-sampled LFBOTs with multi-epoch observations show this behavior: AT\,2018cow \citep{margutti_cow, perley_2019, 2019ApJ...871...73H}, CSS\,161010 \citep{coppejans_2020, gutierrez_2024} and AT\,2024wpp \citep{pursiainen_2025, 2025arXiv250818359O, 2025ApJ...993L...6N, 2025arXiv250900951L, 2026arXiv260103337P}. These objects maintain a smooth blackbody spectrum from 3000 to $9000\,\text{\AA}$ for at least ten days following their discovery, eventually showing H$\alpha$, \ion{He}{1} $\lambda$5876, and \ion{He}{2} $\lambda$4686 emission after several weeks \citep{margutti_cow, gutierrez_2024, 2025arXiv250900951L}. A handful of other LFBOTs show similarly fast-evolving light curves, but have less extensive optical and UV spectral coverage \citep{2020ApJ...895...49H, 2021MNRAS.508.5138P, 2022ApJ...926..112B, 2022ApJ...934..104Y, 2023RNAAS...7..126M, 2023Natur.623..927H, 2024MNRAS.527L..47C, 2024TNSAN.206....1F}. 

The origin of LFBOTs remains an open question, and unifying the extensive multi-wavelength observations under a single model presents a significant challenge. Proposed progenitors are a magnetar-powered SN \citep{2018ApJ...865L...3P}, a failed SN collapsing into a black hole and forming an accretion disk \citep{2019MNRAS.485L..83Q, 2025arXiv251003402C}, a merger of a black hole and Wolf-Rayet star \citep{2022ApJ...932...84M, 2025arXiv251009745K}, or a TDE around a stellar or intermediate mass black hole \citep{perley_2019, 2021ApJ...911..104K, 2025ApJ...986...84T}. Understanding the conditions that drive optical featurelessness could clarify the astrophysical nature of LFBOTs. Notably, a fraction of TDEs also show a prolonged lack of optical and UV line features, at both high and low luminosities \citep[e.g.,][]{hammerstein_2023, 2022ApJ...930...12H, 2023ApJ...955L...6Y, wise_tde, 2025ApJ...989...54H, 2025arXiv251022211Z}, suggesting that similar environments may be present in both featureless TDEs and LFBOTs. 

Several physical processes might generate the optical and UV emission from luminous transient objects. An extended medium, such as a large accretion disk or a quasi-static envelope, could reprocess radiation from a compact central engine to optical wavelengths \citep{roth_2016, 2018ApJ...859L..20D, 2022ApJ...937L..28T}. Expanding optically thick winds could both reprocess a central source of radiation and advect photons outward in the flow \citep{2009MNRAS.400.2070S, 2016MNRAS.461..948M,2020ApJ...894....2P, 2020ApJ...897..156U, 2024arXiv241118985C}. Shocks within the expanding ejecta — driven by fallback accretion or interaction with the surrounding circumstellar medium (CSM) — could release energy, if the debris is dense enough for the radiation to thermalize \citep{2023ApJ...944...74M, 2024ApJ...972..140K, 2025ApJ...986...84T, 2026arXiv260118887G}. How each of these scenarios could produce a long-lived featureless spectrum remains unclear.

Regardless of the luminosity source of these systems, their smooth blackbody continua indicate that the spectra form in an extended, optically thick medium in which the radiation field is thermalized. Under these conditions, we may expect the atomic level populations to approach their local thermodynamic equilibrium (LTE) values. This approximation provides a simple framework to examine the dependence of line emission on source luminosity and ejecta mass, radius, expansion velocity and composition. We explore this parameter space to determine the conditions that produce specific spectral features or maintain featurelessness in luminous transients.

In Section \ref{sec:modeling}, we outline our spectral modeling procedure. Section \ref{sec:spec_form} discusses how the continuum and line profiles form in a scattering-dominated medium. Section \ref{sec:analytics} introduces analytic  estimates that describe the LTE photospheric conditions which produce or suppress conspicuous features. Section \ref{sec:param_survey} presents one-dimensional radiative transfer calculations that explore spectral behavior and line strength across a range of luminosities and gas properties such as ejecta velocity, mass and composition. Section \ref{sec:discussion} connects our analysis to the physical environments of LFBOTs and TDEs. Section \ref{sec:conclusions} summarizes our conclusions. 

\section{Spectral Modeling}\label{sec:modeling}

We model the radiation transport through a spherically symmetric cloud of mass $M$, remaining agnostic about its astrophysical origin. We consider either (i) a single prompt explosion resulting in a freely-expanding, homologous cloud with $v(r) = r/t$, where $t$ is the time since explosion, or (ii) a continuous wind with a constant velocity $v(r) = v_w$. Future work will study emission line shapes for more general velocity profiles. 

The ejecta mass is distributed as a broken power law, 
\begin{align}\label{eqn:density}
    \rho(r) = \begin{cases}
    \zeta_\rho \frac{M}{r_t^3} 
    \left(\frac{r}{r_t}\right)^{-d} {\rm if\,}r < r_t \\
    \zeta_\rho \frac{M}{r_t^3} 
    \left(\frac{r}{r_t}\right)^{-s} {\rm if\,}r > r_t
    \end{cases}
\end{align}
which has a shallow inner region of the ejecta transitioning to a steeper cutoff past the characteristic radius $r_t$. Here $\zeta_\rho$ is a dimensionless constant that depends on the power-law exponents, determined by normalizing the density profile to the total mass. In general, $1 < d < 3$ and $8 < s < 10$; we adopt $d = 2$ and $s = 8$ for our analytic and numerical calculations. We have explored other density profiles and find the emergent spectrum remains largely insensitive to the choice of power-law index; the line strengths depend primarily on the local conditions near the photosphere rather than the global density structure. Our fiducial index of $d = 2$ is also motivated by mass continuity for a spherically symmetric wind \citep{2020ApJ...894....2P}, and is consistent with supernova envelopes \citep{2021MNRAS.502.3385M} as well as TDE disks \citep{2014ApJ...781...82C}.

A steady-state wind described by the case $d=2$ has a density profile
\begin{align}
    \rho(r) = \frac{\dot{M}}{4 \pi r^2 v_w}
\end{align}
with mass-loss rate $\dot{M}$ and wind velocity $v_w$. This relates to our density profile above by $\dot{M} = 4 \pi \zeta_\rho M/t_w$ and $r_t = v_w t_w$, where $t_w$ is the wind expansion time. The density beyond the wind's outer edge $r_t$ depends on the details of the ramp-up of the mass-loss rate, which we model using a steeper power law with index $s = 8$.

We use \texttt{Sedona}, a Monte Carlo radiative transfer code, to perform one-dimensional spectral modeling of transient events \citep{2006ApJ...651..366K}.   
The input source luminosity $L$ is emitted as a blackbody from a reflective spherical core boundary at radius $r_{\rm in}$, with temperature $T_{\rm core}= (L/4\pi \sigma r_{\rm in}^2)^{1/4}$. For most models, we choose $r_{\rm in}$ below the thermalization radius ($r_{\rm th}$) across all wavelengths to ensure the output spectrum is insensitive to the input source. In an optically-thick, freely-expanding cloud, some of the injected radiation is trapped and a portion of this energy is lost to adiabatic expansion. The code automatically adjusts the input luminosity to ensure that $L$ represents the emerging luminosity detected by an observer.

We divide the spatial grid into 256 logarithmically spaced bins between $r_{\rm in}$ and the outer domain boundary at a radius where the optical depth becomes negligible. The frequency grid spans $3 \times 10^{13}$\,Hz to $3.5 \times 10^{17}$\,Hz covering the soft X-ray, UV, optical and infrared (IR) regimes. Our calculations assume steady-state conditions, iterating 25 times to ensure sufficient convergence of the gas temperature profile.

The gas has solar abundances of hydrogen, helium, carbon, nitrogen, oxygen, sulfur, calcium, silicon, iron, nickel and cobalt. To explore different progenitor scenarios, we also study helium-rich models, where the helium fraction by mass is the sum of the solar abundance of H and He, and carbon/oxygen-rich models, where carbon and oxygen each constitute half the solar H and He mass fractions. In both cases, the remaining heavier elements are fixed at their solar abundance ratios.

We focus here on systems that are optically thick, such that at most radii the radiation field is approximately blackbody. In this case, radiative transitions will drive the atomic level populations close to LTE. We therefore assume an LTE ionization and excitation state throughout the material. This approach allows for an extensive survey of parameter space, and is likely to be reasonable for studying the conditions in LFBOTs, which are observed to have blackbody continua. For models with a diffuse medium and a non-thermal spectrum, a non-LTE (NLTE) treatment is needed. We discuss the potential impact of NLTE in \S\ref{sec:discussion} and will explore calculations in this regime in future work.

We treat lines as purely absorptive ($\epsilon$ = 1), while for the continuum, we compute $\epsilon$ at each wavelength based on the local gas conditions. The continuum opacity includes contributions from electron scattering as well as free-free, bound-free, and bound-bound transitions, computed using atomic data from CMFGEN \citep{1998ApJ...496..407H, 2011Ap&SS.336...87H}. The effects of inverse Compton scattering are included, and may become important for broadening line features in high-temperature conditions. We use resolved line transitions, rather than adopting the Sobolev or expansion opacity approximations. This allows for more accurate consideration of the strength and shape of spectral features in non-homologous flows. To reduce computational cost, the lines are artificially Gaussian broadened to have an intrinsic linewidth of 0.2$v_w$. As this is smaller than the outflow velocities, artificial broadening will have minor effects on the line profiles.

\section{Spectrum Formation in a Scattering-Dominated Medium}\label{sec:spec_form}

Before describing the conditions for line formation, we discuss the formation of the underlying continuum in luminous transients. In high temperature conditions, hydrogen is highly ionized, creating a scattering-dominated medium. Continuum radiation forms near the wavelength-dependent thermalization depth, $r_{\rm th}$, below which photons are effectively absorbed and re-emitted.  
We define $r_{\rm th}$ where the radial optical depth $\tau = 1/\sqrt{\epsilon}$, with $\epsilon = \alpha_{\nu}/(\sigma_{\nu} + \alpha_{\nu})$ as the probability a photon of frequency $\nu$ is absorbed by a medium with absorption coefficient $\alpha_{\nu}$ and scattering coefficient $\sigma_{\nu}$. 

Radiation in the optical and UV thermalizes at 
different radii, as $\tau$ and $\epsilon$ depend strongly on wavelength. In the optical, electron-scattering opacity dominates while free-free opacity and bound-free transitions from excited states provide a small absorptive component, with $\epsilon_{\rm opt} \sim 10^{-4}-10^{-3}$. The optical thermalization radius thus lies well beneath the surface where the electron scattering optical depth $\tau_{\rm es} = 1$. In the near UV ($\lambda = 1000 - 3000\,$\AA) the absorptive opacity is high due to low-lying bound-free and bound-bound transitions, such that $\epsilon_{\rm uv} \approx 1$ and the UV optical depth $\tau_{\rm uv} > \tau_{\rm es}$. 

In an expanding medium, photons can be trapped and advected with the flow. The spectrum is Doppler shifted redward while preserving its shape, an effect that can contribute to the reprocessing of radiation to longer wavelengths. Radiation eventually decouples from the flow at the trapping radius, $r_{\rm tr}$, where $\tau \sim c/v$. 

For the power-law density profile of Eq.~\ref{eqn:density}, the optical depth from a coordinate $r > r_t$ to infinity is
\begin{align}
    \tau(r) = \frac{\zeta_\rho}{s-1} \frac{M \kappa}{r_t^2} 
    \left(\frac{r}{r_t}\right)^{-s+1},
    \label{eq:tau}
\end{align} 
and the electron-scattering photosphere ($\tau_{\rm es} = 1$) is 
\begin{align}\label{eqn:phot}
    r_{\rm p} &= r_t 
    \left[ 
    \frac{\zeta_\rho}{s-1} \frac{M \kappa_{\rm es}}{r_t^2} 
    \right]^{\frac{1}{s-1}}.
\end{align}
The thermalization and trapping radii are then
\begin{align}\label{eqn:therm_and_trap}
    r_{\rm th} = \epsilon^{\frac{1}{2(s-1)}} ~r_p 
    ~~~~~~~~~~
    r_{\rm tr} = \left( \frac{v}{c} \right)^{\frac{1}{s-1}} ~r_p.
\end{align}
These radii all scale as $r_t^{(s-3)/(s-1)}$, and for steep density profiles ($s \rightarrow \infty$), they form near $r_t$. Trapping effects will be relevant when
$r_{\rm tr} > r_{\rm th}$, which occurs for expansion velocities $v > c \sqrt{\epsilon} \approx 0.03c$.

\begin{figure*}
    \centering
    \includegraphics[width=\linewidth]{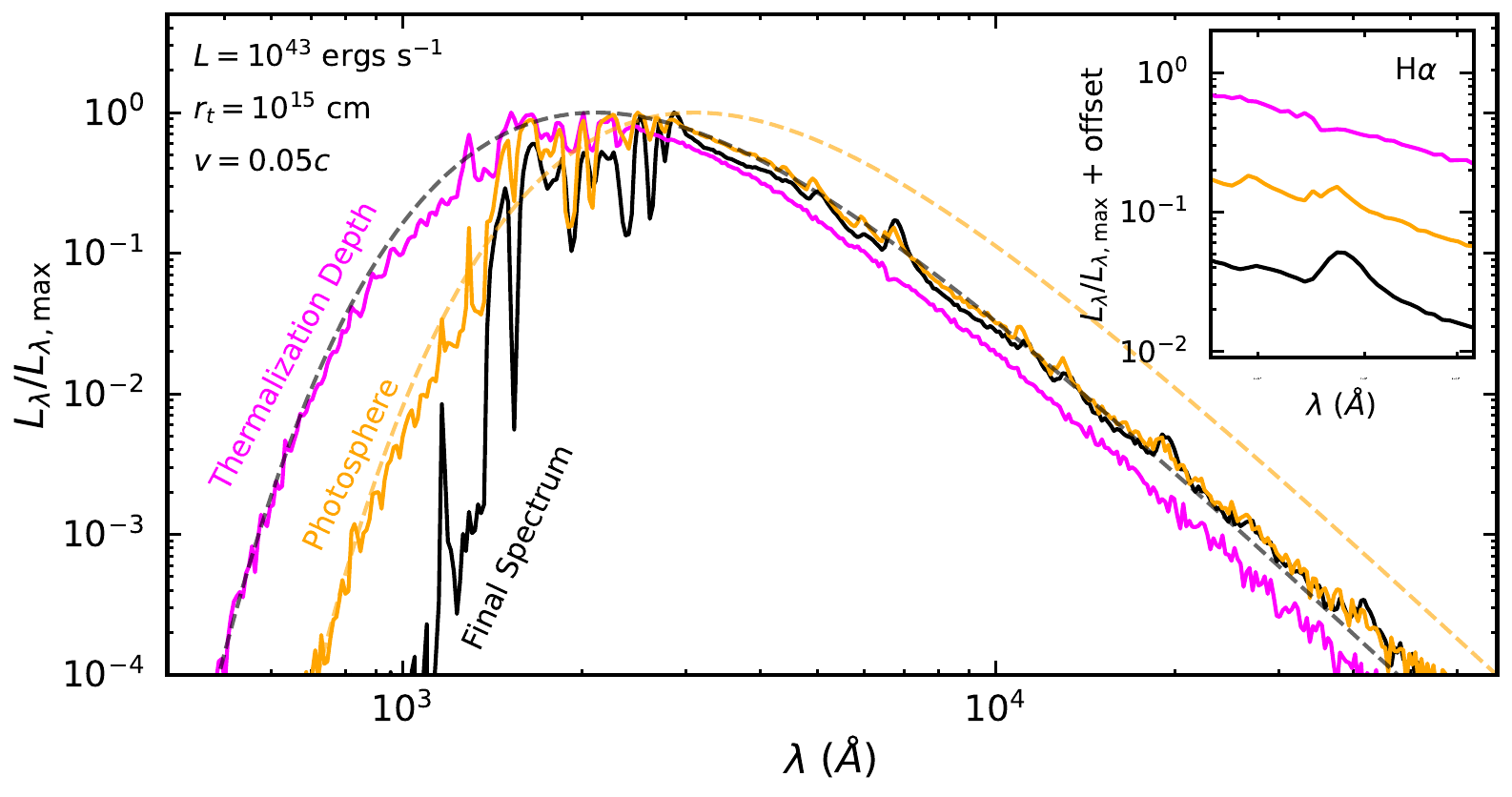}
    \caption{The comoving luminosity at the thermalization depth (magenta) and photosphere (orange) as well as the final outgoing spectrum (black) from a $1 \,M_{\odot}$ cloud with $r_t = 10^{15}\,\rm cm $ surrounding a source with $L=10^{43}\,\rm erg~s^{-1}$. Each radius is calculated in the optical wavelength range (at 6000\,\AA), where electron scattering opacity dominates. The inset shows the H$\alpha$ line emitted at each depth, forming up to the outgoing spectrum. The dashed black curve shows the blackbody fit to the optical band from 3000 to 9000\,\AA, while the dashed orange curve denotes the blackbody produced at the photospheric radius and temperature. The optical continuum is largely produced near $r_{\rm th}$, while emission and absorption features form closer to the photosphere. The blackbody at the photosphere corresponds to the emission in the UV, as the continuum below 3000\,\AA~forms in cooler gas near the edge of the ejecta, and is suppressed relative to the blackbody fit. To mitigate Monte Carlo noise, we re-bin the spectra with 800 logarithmically spaced bins.
    }
    \label{fig:continuum}
\end{figure*}

Figure~\ref{fig:continuum} shows the comoving luminosity at different radii for a model of $1\,M_{\odot}$ ejecta with $r_t = 10^{15}\,\rm cm$, $v = 0.05c$, and $L = 10^{43}\,\rm erg~s^{-1}$,  illustrating how the emergent spectrum develops through the medium.  At the optical thermalization depth, $r_{\rm th}$, the entire spectrum is nearly blackbody at the local temperature, regardless of the spectrum emitted from deeper within the atmosphere. This blackbody is redshifted as radiation is advected to the trapping radius, enhancing the optical continuum. The more opaque UV continuum remains thermalized at and above the $\tau_{\rm es} = 1$ surface, and thus  reflects the cooler temperatures of the outermost gas layers. 

The process of continuum formation may change when the thermalization depth drops deep into the cloud and the photosphere recedes to the bulk of the wind. The UV emission will no longer be effectively reprocessed into the optical when $r_{\rm p} < r_t$, or 
\begin{align}
    M < \frac{r_t^2(s-1)}{\zeta_{\rho}\kappa_{\rm es}}\approx 0.13\,r_{15}^{2}\,\kappa_{0.4}^{-1}~M_{\odot}
\end{align}
for $r_{15} = r_t/10^{15}~\rm cm$ and $\kappa_{0.4} = \kappa_{\rm es}/0.4$. In the above expression, we adopt $d=2$ and $s = 8$, which sets $\zeta_{\rho} = 5/(24\pi)$. In the case of a steady-state wind, this limit can be written as 
\begin{align}
    \dot{M} < \frac{4 \pi v_w^2 t_w (s-1)} {\kappa_{\rm es}}\approx 2.7 \, v_{w, 0.1}^2\,t_{w, \rm d}\,\kappa_{0.4}^{-1} ~M_{\odot}\,\rm yr^{-1}
\end{align}
for $v_{w, 0.1} = v_w/0.1c$ and $t_{w, d} = t_w/1$\,day. In these phases, the emission below 3000\,\AA\, is not reprocessed to longer wavelengths, so the continuum in the optical is low in comparison to the UV. The spectrum is likely to deviate from blackbody, and NLTE effects will become more important. 

Figure~\ref{fig:continuum} shows that, for this model, emission features form near and above the optical photosphere. Line photons produced deeper within the gas are typically reabsorbed in the line, or washed out by repeated electron scatterings. Although the gas in the line-forming region is cooler than the gas at $r_{\rm th}$, features generally form in emission because of the greater volume of the line-emitting region. A similar effect produces emission lines in optically thick stellar winds from hot O, B and Wolf-Rayet stars \citep{2012A&A...544A..67B, 2007ARA&A..45..177C}.

\section{Analytic Conditions for Featureless Spectra}\label{sec:analytics}

In this section, we develop an analytic framework for estimating the strength of line features in LTE. This provides some intuition for the source and ejecta properties that are favorable for suppressing or enhancing spectral features. 

\subsection{Estimating Emission Line Strength}

The line emissivity of a transition from upper level $u$ to the lower level is 
\begin{align} 
    \eta = n_u A \frac{hc}{\lambda_0}.
\end{align}
Here $n_u$ is the number density of atoms in the upper level of the transition, $A$ is the Einstein coefficient and $\lambda_0$ is the center  wavelength of the line. Line features primarily form near the photosphere (Fig.~\ref{fig:continuum}), so we approximate the line emissivity as constant within a shell of thickness $\Delta r$ at radius $r_{p}$. The total line luminosity is then 
\begin{align}\label{eqn:L_line_total}
    L_{\rm line} \approx \eta \times V = n_u A \frac{hc}{\lambda} \times 4 \pi r_{\rm p}^2 \Delta r.
\end{align}
For gas in LTE at temperature $T$, the Boltzmann distribution gives the number density of atoms in excited state $u$,
\begin{align}
n_u = n_i \frac{g}{Z_i} e^{-\Delta E/k_B T} 
\end{align}
where $n_i$ is the number density of atoms in the lower energy state, $g$ is the degeneracy factor of the level, and $\Delta E$ is the excitation energy to level $u$ above the ground state. The Saha equation gives the ratio of $n_i$ relative to the ionization stage above it,
\begin{align}
    \frac{n_{i+1}}{n_i}  =  \Omega~~~~{\rm where} ~\Omega = \frac{2}{n_e \lambda_T^3}
    \frac{Z_{i+1}}{Z_{i}} e^{-\chi/k_B T}, 
     \label{eq:Omega_ion}
\end{align}
and $\chi$ is the ionization energy, $Z_i$ and $Z_{i+1}$ are the partition functions of state $i$ and $i+1$, respectively, $\lambda_T$ is the electron thermal deBroglie wavelength and $n_e$ is the number density of free electrons. If we assume most atoms are in one of these two ionization states, we can write $n_i= n - n_{i+1}$, with $n$ as the total number density of a particular species. We then find $n_i = n/(1 + \Omega)$. Dividing $L_{\rm line}$ by the linewidth $\Delta \lambda$ gives the specific luminosity of the feature,
\begin{align}\label{eqn:L_lambda_line}
    L_{\lambda,\rm line} \approx n \frac{g}{Z_i} \frac{e^{-\Delta E/k_B T}}{1 + \Omega} A \frac{hc}{\lambda} \times \frac{4 \pi r_{\rm p}^2 \Delta r}{\Delta \lambda}.
\end{align}
Here $\Delta \lambda = \lambda_0 \times v/c$, with line center wavelength $\lambda_0$ and line-of-sight velocity $v$. We set $n = n_{\rm p}$, the number density of a given species at $r_{\rm p}$, and $T = T_{\rm p}$, the gas temperature at the photosphere.

To determine when the feature will be detectable against the continuum, we can compare this to the specific continuum luminosity,
\begin{align}
L_{\lambda, \rm cont} = 4 \pi^2 r_{\rm p}^2 B_{\lambda}(T_{\rm p}) \approx  8 \pi^2 r_{\rm p}^2 \frac{c k_{\rm B} T_{\rm p}}{\lambda^4},
\label{eqn:L_lambda_cont}
\end{align}
where the last expression uses the Rayleigh-Jeans approximation. While the Rayleigh-Jeans limit is not strictly valid for some systems with lower gas temperatures, we adopt it here to construct a simplified scaling relation.

The ratio of the line luminosity (Eq.~\ref{eqn:L_lambda_line}) and the continuum luminosity (Eq.~\ref{eqn:L_lambda_cont}) at wavelength $\lambda_0$ is therefore 
\begin{align}\label{eqn:L_ratio}
    \frac{L_{\lambda,\rm line}}{L_{\lambda,\rm cont}} &\approx \frac{g}{Z_i}\frac{Ahc}{2 \pi k_{\rm B}}
    \frac{e^{-\Delta E/k_{\rm B} T_{\rm p}}}{1 + \Omega}
    \frac{n_{\rm p}\Delta r \lambda_0^2}{T_{\rm p} v}. 
\end{align}
In the high-temperature limit relevant to high-energy transients, the medium is mostly ionized, so $1+ \Omega \approx \Omega \propto \frac{T^{3/2}}{n_e} e^{-\chi/k_B T}$ and $n_{\rm p} \approx n_e$. We also approximate $\Delta r \propto r_t$, since all characteristic radii form near $r_t$ for steep density profiles (Eqs.~\ref{eqn:phot} and \ref{eqn:therm_and_trap}). The line-to-continuum ratio simplifies to
\begin{align}\label{eqn:simple_ratio}
    \frac{L_{\lambda,\rm line}}{L_{\lambda,\rm cont}} \propto \frac{n_{\rm p}^2}{T_{\rm p}^{5/2}}\frac{r_t \lambda_0^2}{v} ~e^{\frac{\chi - \Delta E}{k T_{\rm p}}}.
\end{align}
This ratio provides a rough estimate of the visibility of an emission line relative to the continuum. The expression only holds when the species is ionized (i.e., $\Omega \gtrsim 1$ in Eq.~\ref{eq:Omega_ion}), which occurs in hydrogen when the temperature is greater than
\begin{align}
k T_{\rm ion} \approx \frac{\chi}{\zeta} \approx 
6000~\frac{\chi_{13.6}}{\zeta_{25}} ~{\rm K},
\end{align}
where $\zeta = \log(2 Z_{1+1}/Z_i n_e \lambda_T^3) $, $\chi_{13.6} = \chi/13.6$~eV and $\zeta_{25} = \zeta/25$.

\begin{figure*}
    \centering
    \includegraphics[width=\linewidth]{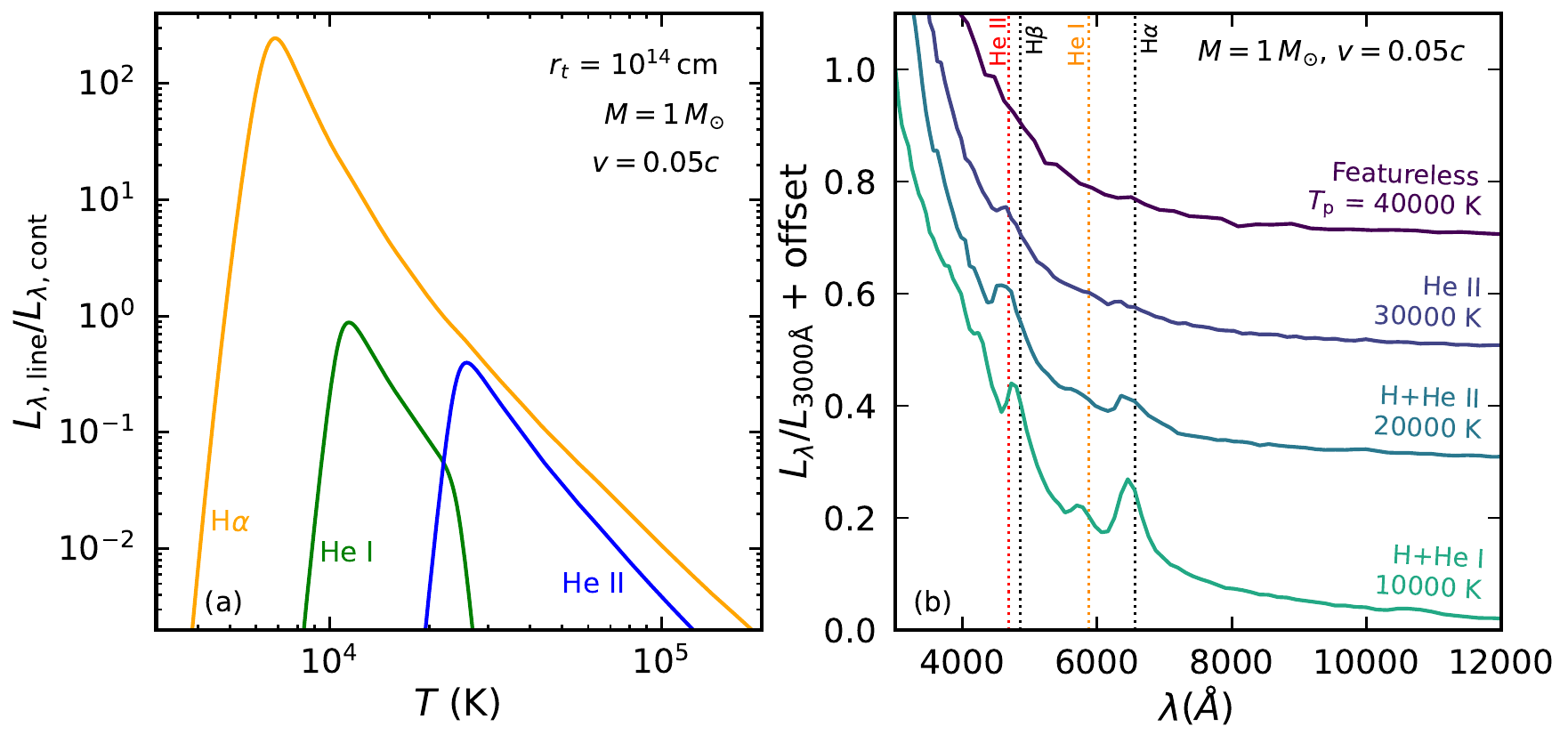}
    \caption{(a) Estimate of the line-to-continuum luminosity ratio (Eqs.~\ref{eqn:L_lambda_line} and \ref{eqn:L_lambda_cont}) for H$\alpha$ (orange), \ion{He}{1} $\lambda$5876 (green) and \ion{He}{2} $\lambda$4686 (blue) as a function of $T$, evaluated at the photosphere. We take a $1\,M_{\odot}$ cloud composed of H and He with a power-law density structure and $r_t=10^{14}\,\rm cm$. We expect features to evolve as the temperature varies. (b) Synthetic spectra of models with photospheric temperatures in the range 10000 to 40000 K. We use the broken power law density profile with a $1\,M_{\odot}$ ejecta and an expansion velocity of $0.05c$. We start with $L = 10^{42}\,\rm erg \,s^{-1}$ and $r_t = 2.8 \times 10^{14}$\,cm for the $T_{\rm p}= 10000$\,K model and progressively increase $L$ and decrease $r_t$ up to  $L = 10^{44}\,\rm erg \,s^{-1}$ and $r_t = 9 \times 10^{13}\, \rm cm$ to reach the $T_{\rm p} = 40000$\,K model. As the temperature goes up, the ionization state of the gas increases and emission features weaken; we see the spectra transition from showing Balmer lines to only helium features, until complete line suppression is achieved at high $T_{\rm p}$. 
    }
    \label{fig:T_series}
\end{figure*}

Figure~\ref{fig:T_series}a shows the value of $L_{\lambda,\rm line}/L_{\lambda,\rm cont}$ (Eq.~\ref{eqn:L_ratio}) at varying $T$ for H$\alpha$, \ion{He}{1} $\lambda$5876 and \ion{He}{2} $\lambda$4686. We calculate the ratio by iteratively solving the Saha distribution for the LTE level populations of each species in a $1~M_{\odot}$ gas cloud composed of hydrogen and helium, with $r_t = 10^{14}~\rm cm$ and $v = 0.05c$. The continuum level is set by the full Planck function, and we take $\Delta r = r_{\rm p} - r_{\rm th}$, the size of the line-forming region. We assume that $\eta$ is constant over the thin emitting shell, that reabsorption effects are negligible and that features form in gas at a single temperature. Initially, line emission rises exponentially with $T$ as electrons gain sufficient energy to populate the excited states, but starts to weaken as the species ionizes above $T_{\rm ion}$. For large enough temperature, the line-to-continuum ratio decreases as $T^{-5/2}$ (Eq.~\ref{eqn:simple_ratio}). \ion{He}{1} $\lambda$5876 and \ion{He}{2} $\lambda$4686 emission occur at higher temperatures due to their larger excitation and ionization energies; the spectrum is expected to become featureless under the highest gas temperatures, at which the medium is completely ionized.

Although Figure~\ref{fig:T_series}a presents an approximation of the line-to-continuum ratio of hydrogen and helium features with evolving temperature, Figure~\ref{fig:T_series}b qualitatively supports this spectral sequence with emission from a series of $1\,M_{\odot}$ models with varying $T_{\rm p}$. The highest-temperature gas emits a smooth blackbody spectrum. He-dominated spectra originate from intermediate-$T$ gas, while Balmer lines and \ion{He}{1} appear in cooler gas. Thus, each temperature approximately maps to a different spectroscopic class. 

\subsection{The Parameter Space of Line Emission and Suppression}

We can express Equation \ref{eqn:L_ratio} in terms of the source luminosity $L$ and ejecta properties $M$, $\kappa_{\rm es}$ and $r_t$ using the approximate conditions at the photosphere,
\begin{align}\label{eqn:T_phot}
    T_{\rm p} &= \left( \frac{L}{4 \pi \sigma} \right)^{1/4} \left[\frac{\zeta_{\rho}}{s-1} \frac{M \kappa_{\rm es}}{r_t^2}  \right]^{-\frac{1}{2(s-1)}} r_t^{-1/2},
\end{align}
from $L = 4 \pi \sigma r_{\rm p}^2 T_{\rm p}^4$, and
\begin{align}\label{eqn:n_phot}
    n_{\rm p} &= \frac{\zeta_{\rho}}{\mu m_p} \frac{M}{r_t^3} \left[\frac{\zeta_{\rho}}{s-1} \frac{M \kappa_{\rm es}}{r_t^2}  \right]^{-\frac{s}{s-1}}.
\end{align}
These give
\begin{align}\label{eqn:L_simple}
    \frac{L_{\lambda,\rm line}}{L_{\lambda,\rm cont}}
    &\propto L^{-\frac{5}{8}}~ r_t^{\frac{(s-3)}{4(s-1)}}~ M^{\frac{1}{4(s-1)}} ~v^{-1} ~\lambda_0^{2}~ e^{\xi}
\end{align}
where
\begin{align}
    \xi =  \frac{\chi - \Delta E}{k T_p} \propto L^{-1/4}~r_t^{\frac{s-3}{2(s-1)}} ~M^{-\frac{1}{2(s-1)}}.
\end{align}
In the limit of a very steep density profile ($s \rightarrow \infty$) the ratio scales as 
\begin{align}
\frac{L_{\lambda, \rm line}}{L_{\lambda, \rm cont}} \propto L_{45}^{-\frac{5}{8}} ~r_{15}^{\frac{1}{4}} ~v_{0.05c}^{-1} ~e^{\xi}.
\end{align}
The source luminosity, $L$, and ejecta radius, $r_t$, influence the appearance of emission features by regulating the temperature and therefore ionization state of the gas. Higher luminosities and more compact ejecta produce hotter conditions that suppress emission features. The gas expansion velocity, $v$, broadens lines such that they appear weaker in comparison to the continuum. 

\begin{figure}
    \centering
    \includegraphics[width=\linewidth]{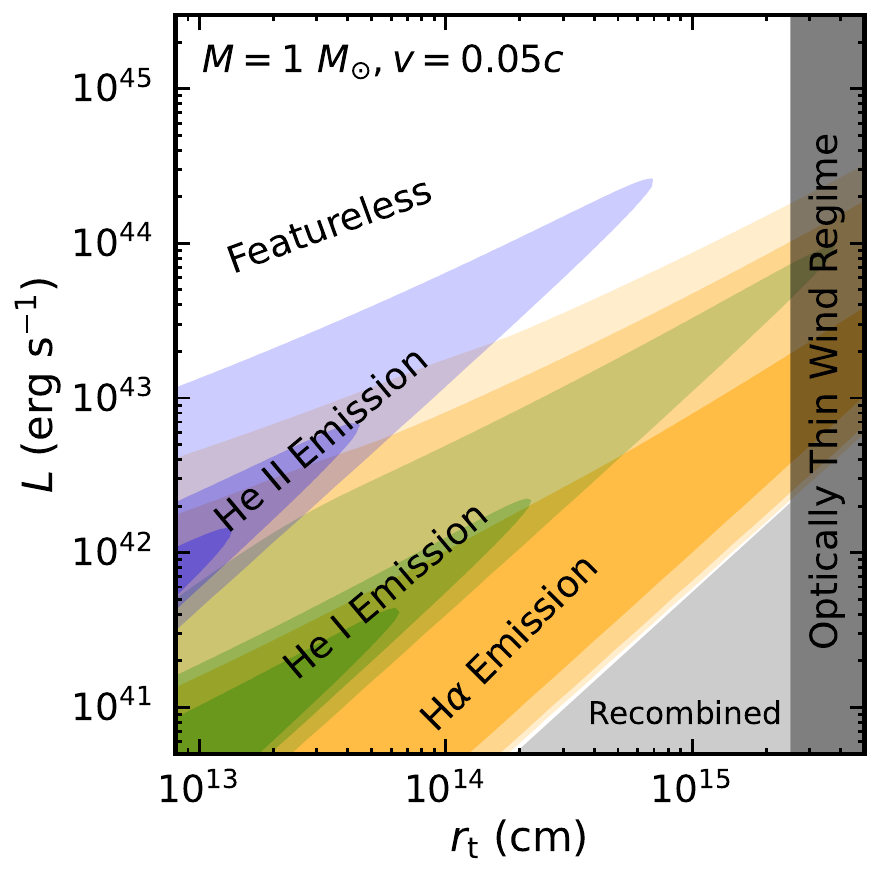}
    \caption{ 
    The parameter space of line emission and suppression as a function of bolometric luminosity, $L$, and ejecta radius, $r_t$, for a $1\,M_{\odot}$ cloud expanding at $v = 0.05c$.  
    The shaded regions show the ratio of line to continuum emission (Eqs.~\ref{eqn:L_lambda_line} and \ref{eqn:L_lambda_cont}), evaluated by iteratively solving the Saha-Boltzmann distribution for the level populations and using the full Planck function for the continuum level. The orange regions represent the conditions where H$\alpha$ appears, with contours for $L_{ \lambda, \rm line}/L_{\lambda, \rm cont}$ $>$ 0.5, 1 and 10. The green contours correspond to \ion{He}{1} $\lambda$5876 emission and the blue contours to \ion{He}{2} $\lambda$4686 emission, with levels at $L_{ \lambda, \rm line}/L_{\lambda, \rm cont}$ $>$ 0.1, 0.5 and 1. For other velocities, the line strength ratio should scale as $v^{-1}$ (Eq.~\ref{eqn:L_simple}). The shaded light gray region to the bottom right indicates the regime where hydrogen has recombined and absorption features are likely to form.  We consider conditions where the photosphere forms near the edge of the ejecta; for $r_t > 2.5 \times 10^{15}\rm \,cm$ (dark gray), the photosphere recedes into the bulk of the outflow, and NLTE effects are likely to become important. }\label{fig:space_with_contours}
\end{figure}

Figure~\ref{fig:space_with_contours} shows the LTE ratio of $L_{\lambda, \rm line}/L_{\lambda, \rm cont}$ from a $1\,M_{\odot}$ ejecta expanding at $v = 0.05c$ across a range of $L$ and $r_t$. In this plot, we solve Equation~\ref{eqn:L_ratio} exactly, including all constants, setting the continuum level with the Planck function and determining the level populations by iteratively solving the Saha distribution in a medium composed of hydrogen and helium. These calculations assume line and continuum formation occur between the photosphere and the thermalization depth in the steep region of the density profile near the edge of the ejecta, with $\Delta r = r_{\rm p} - r_{\rm th}$.

For the regions of high luminosities ($L > 10^{44}$\,erg~s$^{-1}$) and compact radii ($r_t < 10^{14}$\,cm) in Figure~\ref{fig:space_with_contours}, the temperature and ionization is high, resulting in a nearly featureless spectrum. \ion{He}{2} lines become noticeable under intermediate conditions for which helium is in a singly-ionized state. Lower luminosities and larger clouds contain cooler gas, so a sufficient fraction of neutral hydrogen and helium form to produce H$\alpha$ and \ion{He}{1} $\lambda$5876 emission. 

The mapping of line emission shown in Figure~\ref{fig:space_with_contours} depends on the velocity of the medium. Decreasing $v$ increases the prominence of a line feature relative to the continuum (Eq.~\ref{eqn:L_simple}). For lower $v$, the H$\alpha$, \ion{He}{1} and \ion{He}{2} line emission contours would shift upward to cover more of the parameter space. 

Figure~\ref{fig:space_with_contours} is not expected to change much with mass, given the weak dependence (Eq.~\ref{eqn:L_simple}). However, if the mass is low enough such that the photosphere recedes into the bulk of the ejecta and thermalization becomes inefficient, we expect the optical continuum level to drop and the strength of features to increase.  For a $1\,M_{\odot}$ cloud, this occurs for $r_t \gtrsim 2.5 \times 10^{15}\,\rm cm$. When the radiation does not thermalize, our LTE assumption likely breaks down; we therefore limit our study to conditions where the photosphere lies near the outer edge of the ejecta and there is no additional low density circumstellar material surrounding it.

Given the approximations made, Figure~\ref{fig:space_with_contours} is only a suggestive overview of landscape of line emission and suppression in luminous sources. However, the qualitative behavior of spectra for changing gas temperature is confirmed in our spectral models (Fig.~\ref{fig:T_series}); in the following section, we will verify the dependence of line strength on specific source and ejecta properties with calculations of synthetic spectra.

\section{Dependence of Spectral Lines on Matter Properties}\label{sec:param_survey}

We now show with numerical spectral calculations that the analytic results in \S\ref{sec:analytics} reasonably capture the conditions that produce or suppress optical features. We survey models with $L = 10^{41}$ to $10^{45}\, \rm erg~s^{-1}$, $r_t = 5 \times 10^{13}$ to $10^{16}\,\rm cm$, $v = 0.03$ to $0.2c$ and $M = 0.01$ to $10\,M_{\odot}$, as well as solar, He-rich and C/O-rich compositions. For most models, we assume the expansion velocity is constant with radius. We have examined homologous models, and found that the resulting spectra differ from our constant velocity models by less than 10\% in the optical, with minor  variations in the strength and shape of the line features. We defer a more detailed study of velocity structures and emission line profiles for later work.

\subsection{Luminosity}

\begin{figure*}
    \centering
    \includegraphics[width=\linewidth]{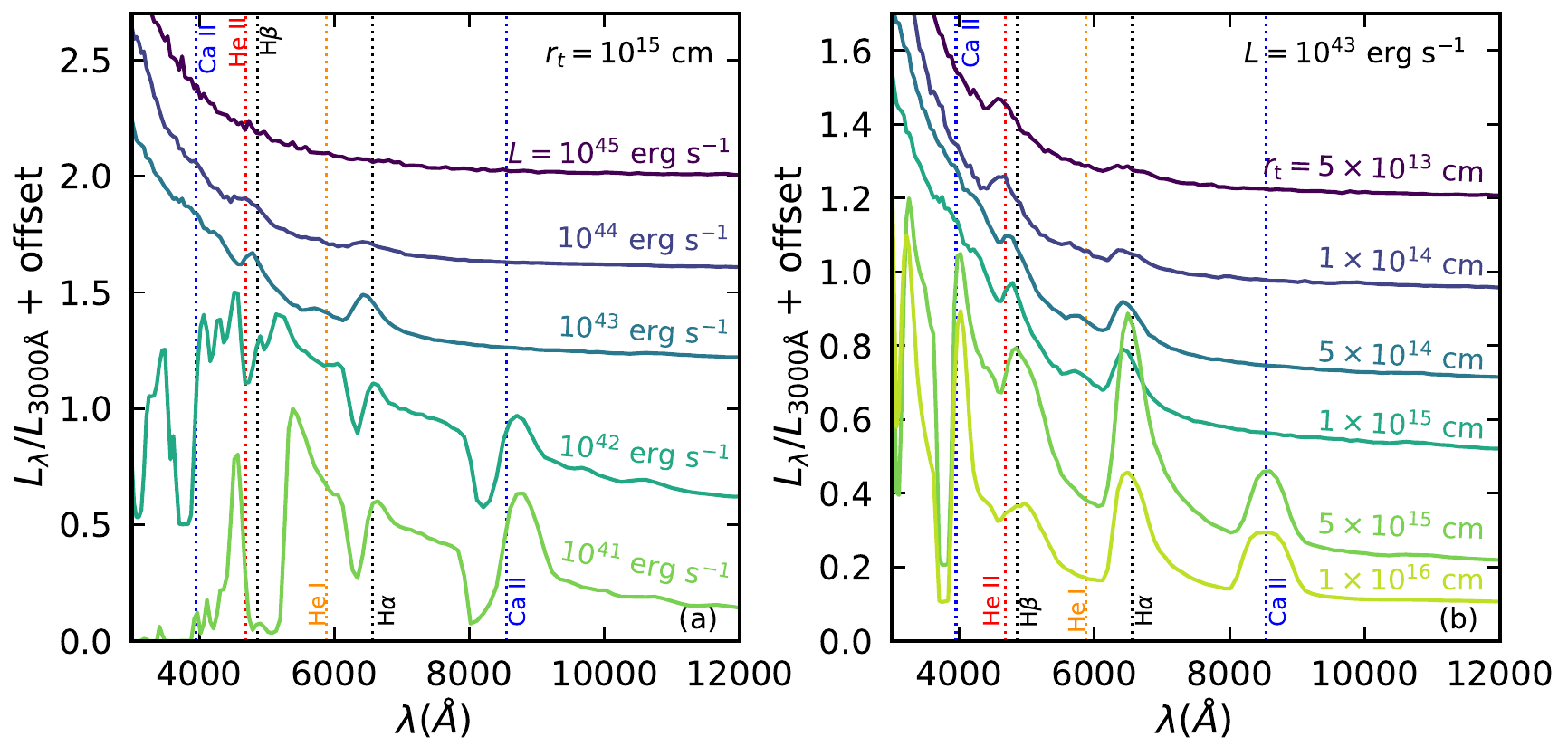}
    \caption{(a) Synthetic optical spectra emitted by $1\,M_{\odot}$ ejecta surrounding sources of five different bolometric luminosities. Each cloud has a solar composition and expands at a constant velocity of $0.05c$. (b) Synthetic spectra from clouds with five characteristic ejecta radii. Each model has a bolometric luminosity of $10^{43}$\,erg~s$^{-1}$, a mass of $1\,M_{\odot}$ and an expansion velocity of $0.05c$. As the luminosity goes up, and as the size of the ejecta shrinks, H$\alpha$ and helium line emission weakens. 
    }
    \label{fig:r_and_L}
\end{figure*}

Figure~\ref{fig:r_and_L}a shows spectra from sources spanning five orders of magnitude in luminosity, from $10^{41}$ to $10^{45}$\,erg~s$^{-1}$. 
A spectral sequence emerges: as the radiation field weakens, line emission strengthens. The highest bolometric luminosity produces a featureless spectrum; under an extreme radiation field, the ejecta reaches high temperatures and is therefore highly ionized, preventing the medium from emitting line photons. At $10^{44}$\,erg~s$^{-1}$, H$\alpha$ and \ion{He}{2} $\lambda$ 4686~\AA\ emission lines appear. Bolometric luminosities of $10^{43}\,\rm erg~s^{-1}$ are insufficient to ionize helium, so \ion{He}{2} lines disappear and weak \ion{He}{1} and H$\beta$ become visible. 

At $L  \lesssim  10^{43}\,\rm erg~s^{-1}$, a sharp hydrogen recombination front forms, above which the atmosphere is optically thin to electron scattering. The photosphere coincides with the front, enabling the formation of typical P-Cygni line features, including \ion{Fe}{2} lines and the \ion{Ca}{2} IR triplet. In this regime, the spectrum resembles a Type IIP SN in the plateau phase. 

\subsection{Ejecta Radius}

Figure~\ref{fig:r_and_L}b shows spectra from atmospheres with five different characteristic radii from $5 \times 10^{13}$ to $10^{16}$\,cm. Another spectral sequence appears as the radius changes: larger clouds emit stronger line features.  While the luminosity of this model is modest ($L = 10^{43}$\,erg~s$^{-1}$), a nearly featureless spectrum may nonetheless result if the ejecta is compact, leading to high temperatures and ionization. For the most compact model ($r_t = 5 \times 10^{13}\,\rm cm$) only weak H$\alpha$ and \ion{He}{2} $\lambda4686$ lines appear in the spectrum. For $r_t = 10^{14}$\,cm, the gas is cool enough that helium recombines and the \ion{He}{2} $\lambda$4686 line disappears, while H$\alpha$, H$\beta$, and \ion{He}{1} line $\lambda$5876 are stronger. The most extended clouds produce the strongest Balmer line emission. 

For our adopted value $s = 8$ and fixed mass, $r_{\rm p} \propto r_t^{5/7}$; the electron scattering photosphere moves outward as the size of the cloud grows. Eventually, the outer layers will grow optically thin and the photosphere will recede into the shallow portion of the ejecta. In this regime, where the density follows $s = 2$, $r_{\rm p} \propto r_t^{-1}$ and the photosphere continues to shrink as the cloud expands. We expect NLTE effects to become important as the bulk of the ejecta becomes optically thin in the continuum, and the cool, extended gas above the photosphere forms conspicuous line features.

\subsection{Velocity}

 \begin{figure}
    \centering
    \includegraphics[width=\linewidth]{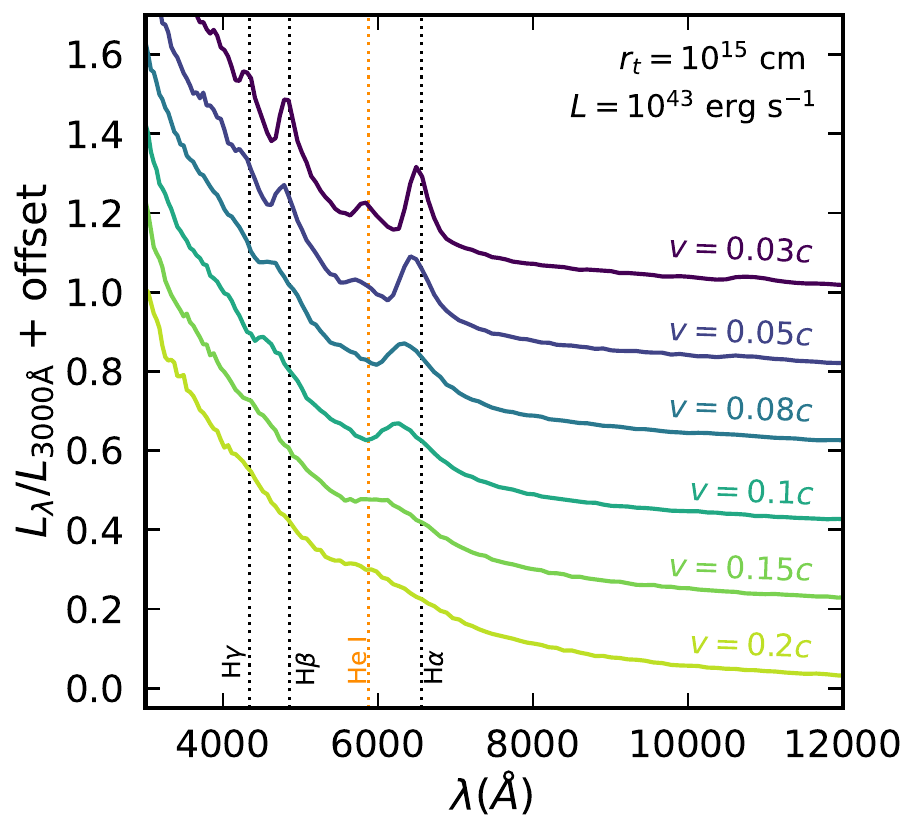}
    \caption{Synthetic spectra from material moving at $0.03c$ up to $0.2c$. Each model has a mass of $1\,M_{\odot}$, a radius of $10^{15}$\,cm and a luminosity of $10^{43}$\,erg~s$^{-1}$. The increased line broadening at higher velocities washes out the emission features and makes them less conspicuous.}
    \label{fig:velocity_series}
\end{figure}

Figure~\ref{fig:velocity_series} shows how varying the ejecta expansion velocity affects spectral features. For these model parameters, the slowest-moving model ($v = 0.03c$)  forms prominent, narrow Balmer lines and \ion{He}{1} $\lambda$5876. As we increase the velocity, line features broaden and become less noticeable relative to the continuum. \ion{He}{1} and H$\gamma$ blend into the continuum at velocities greater than $0.03c$. At $v = 0.1c$, only H$\alpha$ remains visible. 

Doppler broadening can therefore suppress optical features in fast-moving ejecta. However, all of the models in Figure~\ref{fig:velocity_series} are optically thick with the photosphere forming near the edge of the cloud. In a time-dependent scenario, rapidly expanding material may quickly become optically thin and eventually fail to effectively reprocess UV emission to the optical band. We consider these conditions in the following subsection. 

\begin{figure*}
    \centering
    \includegraphics[width=0.9\linewidth]{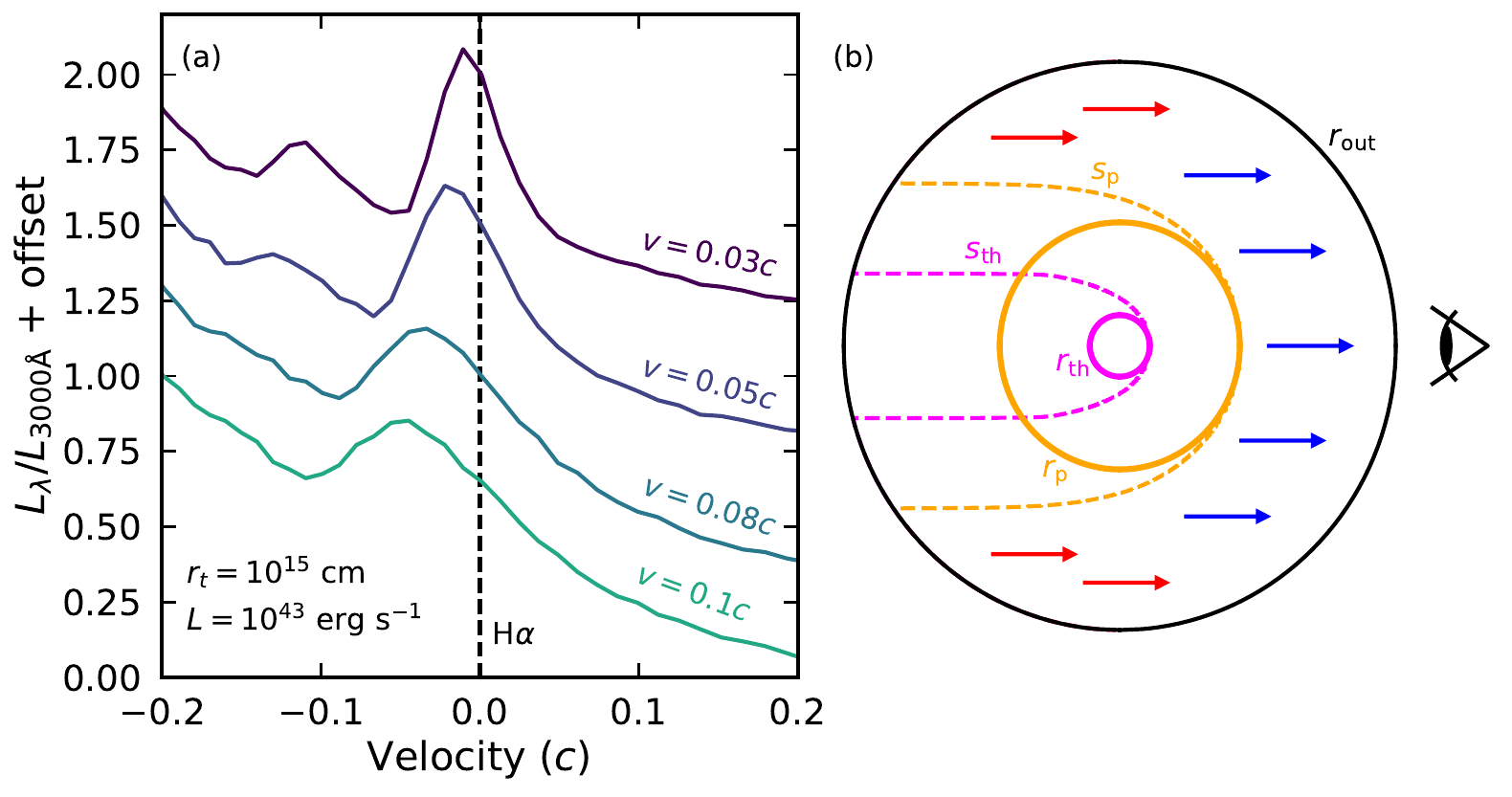}
    \caption{(a) Synthetic H$\alpha$ line profiles from Figure~\ref{fig:velocity_series}, from ejecta with expansion velocities from $0.03c$ to $0.1c$. The line peaks are systematically blueshifted from the rest wavelength (dashed line) by amount proportional to the expansion velocity. (b) A schematic illustrating the origin of the blueshift, with the observer on the right side of the cloud with radius $r_{\rm out}$. The dashed lines represent the surfaces where the optical depth integrated along the observer line of sight is equal to one ($s_{\rm p}$, magenta) and $1/\sqrt{\epsilon}$ ($s_{\rm th}$, orange).  The thermalization sphere ($r_{\rm th}$, magenta) and photosphere ($r_{\rm ph}$, orange) are also marked for comparison. Photons originating from the receding far side of the ejecta (red arrows) are redshifted, while those emerging from the approaching near side are blueshifted (blue arrows). Only photons emitted outside of the $s_{\rm p}$ surface reach the observer without further interaction, and the occlusion of the far side results in the observed line profile being systematically shifted to the blue. }
    \label{fig:velocity_profiles}
\end{figure*}

Figure~\ref{fig:velocity_profiles}a zooms in on the H$\alpha$ line profiles of the four $1\,M_{\odot}$ clouds expanding at velocities from $0.03$ to $0.1c$ shown in Figure~\ref{fig:velocity_series}. The peak of the H$\alpha$ line profile has a systematic blueshift that scales with the expansion velocity of the ejecta by approximately $v/2$. Figure~\ref{fig:velocity_profiles}b illustrates that this blueshift arises because the observer primarily detects photons from approaching gas. The $\tau_{\rm es} = 1$ depth integrated along the line of sight forms the curved surface shown ($s_{\rm p}$). The emitting volume of the cloud behind this surface is occluded by the dense central region of ejecta. While some redshifted emission from receding material near the poles is visible, the photons that do reach the viewer are predominantly blueshifted. The final line profile therefore blueshifts by a fraction of the total expansion velocity.

Scattering off of hot electrons can also broaden line features in luminous transients \citep[e.g.,][]{2018ApJ...855...54R}. This effect will become comparable to the Doppler broadening from outward ejecta expansion when $\tau_{\rm es} \sqrt{k T/ m_e} \gtrsim  v $ or 
\begin{align}
    T \gtrsim \frac{m_ev^2}{k_b \tau_{\rm es}^2} \approx 1.4 \times 10^7\,v_{05}^2 \tau_{\rm es}^{-2}\, \rm K
\end{align}
with electron mass $m_e$, expansion velocity $v_{05} = v/0.05c$ and electron scattering depth $\tau_{\rm es} \approx 1$, as we find lines generally form near the photosphere. Given the typical temperatures near the photosphere in these models ($T_{\rm p}\sim 10^4  - 10^5$~K) the effect of non-coherent scattering on the widths of line features is small compared to Doppler broadening unless the expansion velocity is low ($v \lesssim 0.01c$). At these very low bulk outflow speeds, thermal electron scattering broadens the lines with a characteristic peaked, Lorentzian profile. Doppler shifts from the bulk velocity of the electrons can also contribute to the overall blueshift of the line profile. 

The models in Figures~\ref{fig:velocity_series} and~\ref{fig:velocity_profiles} are constant velocity winds, but we find qualitatively similar linewidths and systematic blueshifts in homologous models with $v \propto r$. The exact  shape of the line profiles will depend on the velocity structure and may differ among expanding, rotating, turbulent, or infalling gas. We defer a study of the detailed line shapes — and the insight they lend into the kinematics of the gas — for later work.  

\subsection{Mass}

 \begin{figure*}
    \centering
    \includegraphics[width=\linewidth]{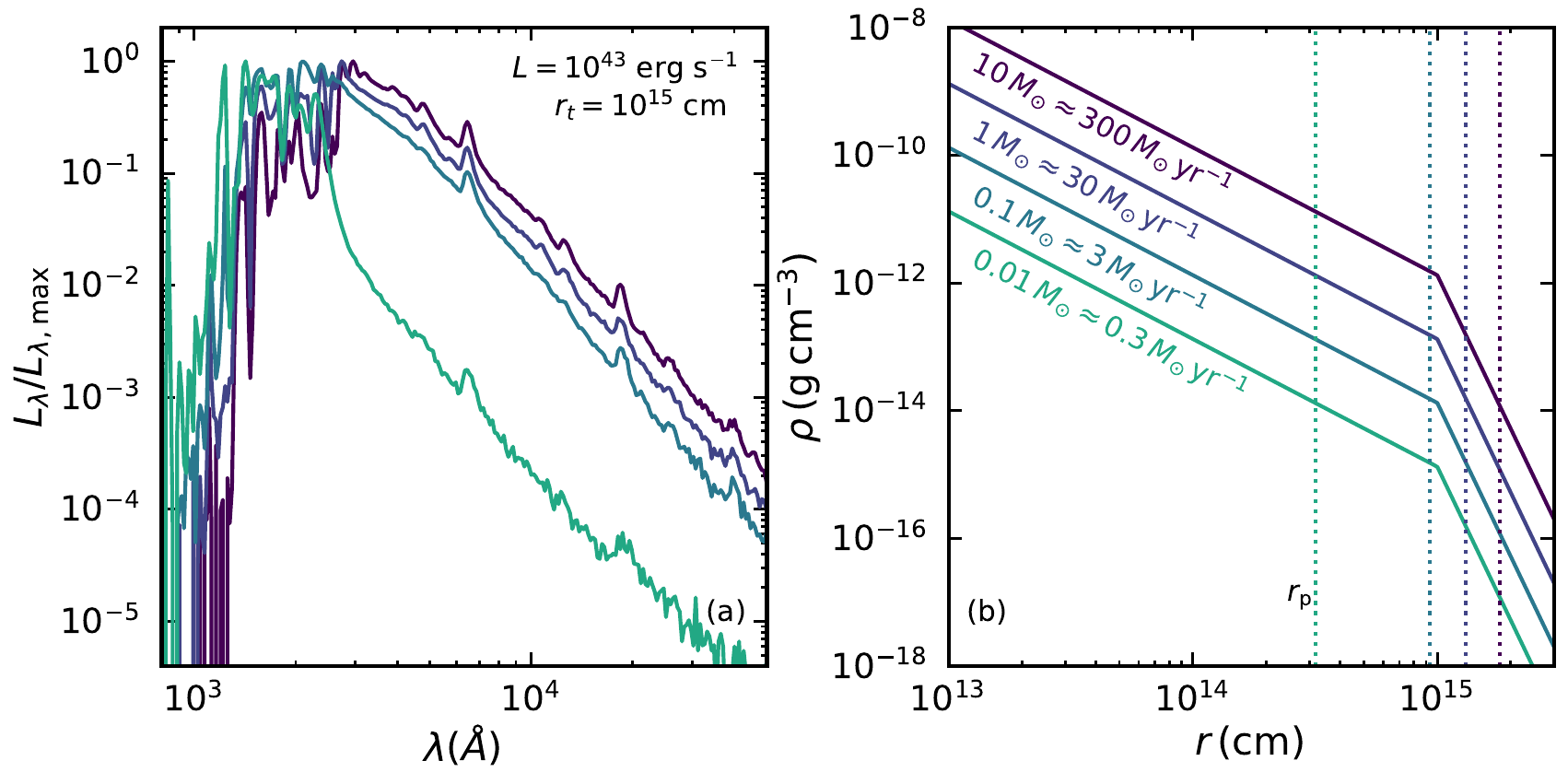}
    \caption{(a) Synthetic spectra produced by four clouds with varying masses from $0.01$ to $10\,M_{\odot}$. Each cloud has a characteristic radius of $10^{15}$\,cm as well as an expansion velocity of $0.05c$, and surrounds a source with a bolometric luminosity of $10^{43}$\,erg~s$^{-1}$. (b) The density profiles for each cloud. We calculate the mass-loss rates corresponding to each atmosphere, and indicate the photospheres of each density profile with the dashed vertical lines. For the $0.01 M_{\odot}$ ($\dot{M} = 0.3\,M_{\odot}\,\rm yr^{-1}$) model, the mass is low enough that the photosphere has receded into the bulk, so the optical continuum is not thermalized and the UV emission is not efficiently reprocessed to optical wavelengths. As a result, the continuum level is suppressed from 3000 to 9000\,\AA, while the UV emission remains bright. 
    }
    \label{fig:mass_series}
\end{figure*}

Figure~\ref{fig:mass_series} shows spectra for models with varying ejecta mass but the same $r_t = 10^{15}$\,cm and $L = 10^{43}$\,erg~s$^{-1}$. Consistent with our analytic expectations (Eq.~\ref{eqn:L_simple}), there is little change in the strength of H$\alpha$ as the mass increases from $0.1$ to $10\, M_{\odot}$. In these models, the ejecta is sufficiently optically thick that the photosphere forms near the edge of the ejecta, regardless of the interior mass, and the physical conditions in the line-forming region are set primarily by the luminosity and outer radius. 

The formation of the optical continuum through thermalization and adiabatic processes occurs throughout the ejecta; the efficiency of reprocessing the UV radiation to the optical decreases somewhat with declining mass. Once the ejecta mass drops to 0.01$\,M_{\odot}$, the thermalization and trapping radii recede to near zero, and the material reprocesses almost none of the UV emission to optical wavelengths. The continuum from 3000 to 9000\,\AA\, drops dramatically, while the UV band remains bright. In order to maintain strong optical emission, the ejecta must remain optically thick enough to keep the photosphere at the edge of the wind.

For ejecta masses that are insufficient to thermalize the optical continuum, the blackbody fit to emission from 3000 to 9000\,\AA\, may underestimate the flux in the UV band. This in turn biases the bolometric luminosity derived from the continuum fit.

\subsection{Composition}

\begin{figure*}
    \centering
    \includegraphics[width=\linewidth]{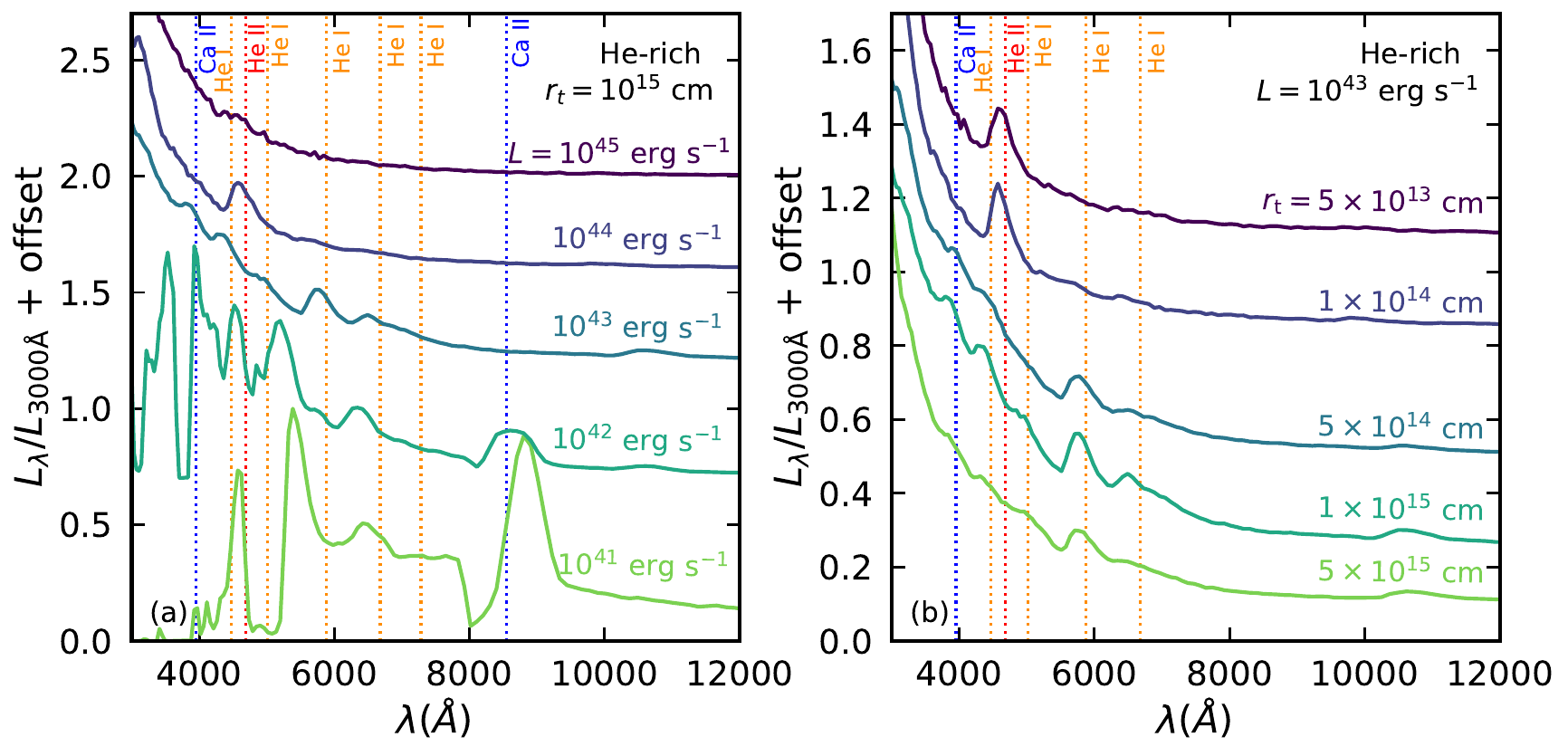}
    \caption{(a) Synthetic optical spectra from helium-rich, $1\,M_{\odot}$ ejecta with five bolometric luminosities. The medium expands at a constant velocity of 0.05$c$. (b) Spectra from helium-rich material with five characteristic radii. Each cloud surrounds a $10^{43}$\,erg~s$^{-1}$ source, has a mass of $1\,M_{\odot}$ and expands at $0.05c$. At the highest luminosities, optical line emission is mostly suppressed. The spectra from compact systems ($r_t < 10^{14}\,\rm cm$) are also mostly featureless, except for an \ion{He}{2} line at 4686\,\AA. 
    }
    \label{fig:r_and_L_He}
\end{figure*}

\begin{figure*}
    \centering
    \includegraphics[width=\linewidth]{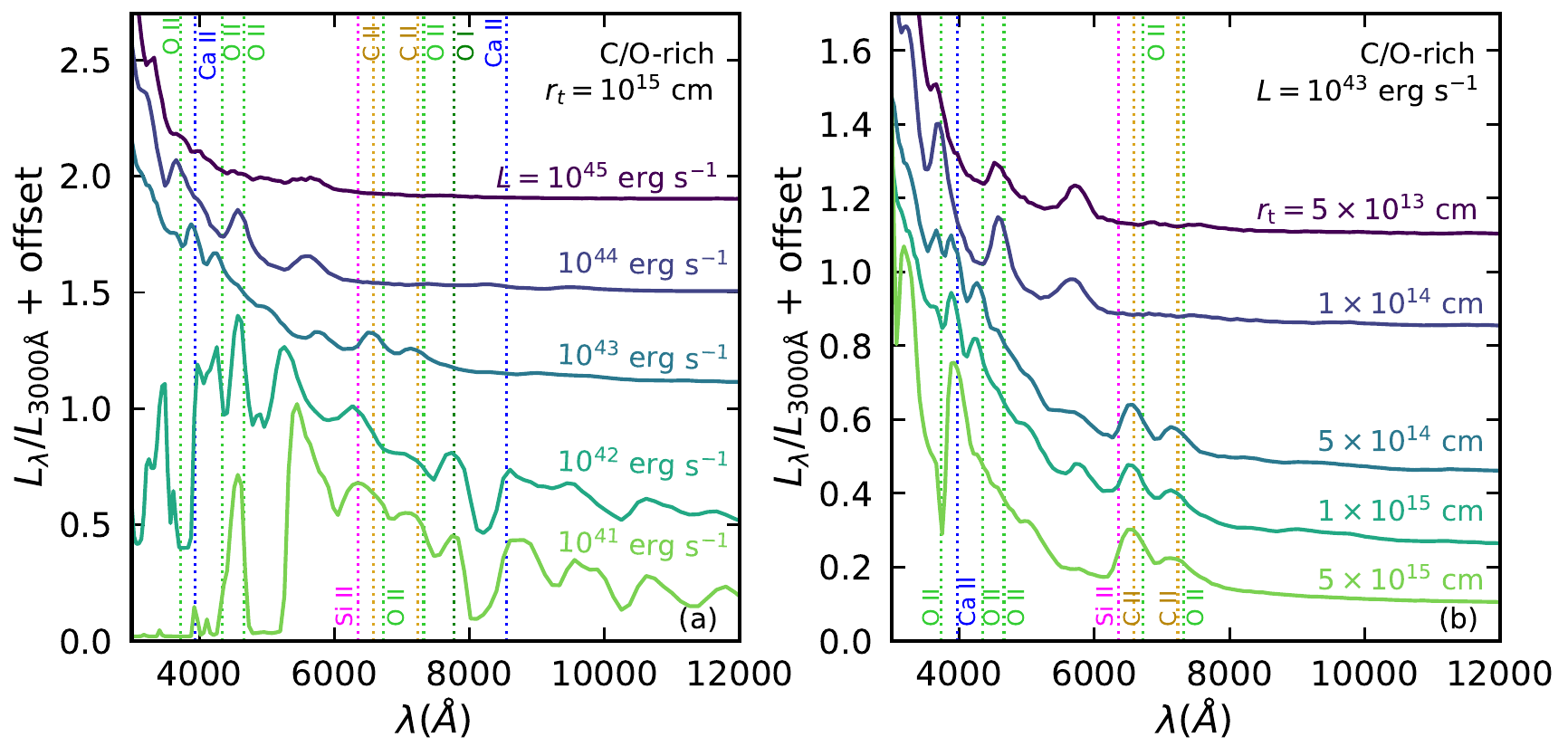}
    \caption{(a) Synthetic optical spectra from carbon/oxygen-rich ejecta with five input bolometric luminosities. Each cloud has $M = 1 M_{\odot}$, $r_t = 10^{15}\,\rm cm$ and a constant velocity of 0.05$c$. (b) Synthetic spectra from carbon/oxygen-rich gas with five characteristic ejecta radii, $M = 1\,M_{\odot}$ and $v = 0.05c$, surrounding a central source emitting $10^{43}$\,erg~s$^{-1}$. While the emission and absorption lines weaken as the source brightens and the cloud contracts, none of the spectra are entirely featureless.}
    \label{fig:r_and_L_CO}
\end{figure*}

Figure~\ref{fig:r_and_L_He}a shows the spectral sequence across source luminosities for a medium composed of 99\% helium and a solar distribution of metals. We see a similar evolution in the spectral sequence for He-rich ejecta as we do in solar-metallicity material, although with more prominent helium lines. 
Figure~\ref{fig:r_and_L_He}b shows the spectral evolution across radii from $5 \times 10^{13}$ to $5 \times 10^{15}\,\rm cm$. Compact systems show a strong \ion{He}{2} emission feature at $4686\,\text{\AA}$; the temperature is high enough to keep helium singly ionized. \ion{He}{2} disappears and \ion{He}{1} features form in spectra from larger clouds, as the material cools and recombines. Thus, the \ion{He}{2} line is only suppressed under the most extreme conditions. 

Figure~\ref{fig:r_and_L_CO} shows spectra from C/O-rich material across the luminosity and radius spectral sequence. At high luminosities ($L > 10^{44}$\,erg~s$^{-1}$), C/O-rich ejecta produces some weak line emission (Fig.~\ref{fig:r_and_L_CO}a). Under intermediate source luminosities, the spectrum becomes highly featured with potential carbon, oxygen and calcium emission and absorption lines. Even lower luminosities, $10^{41}$ and $10^{42}\,\rm erg~s^{-1}$, show iron transitions between 3000 and $5000\,\text{\AA}$. Decreasing the ejecta radius also does little to suppress conspicuous optical lines (Fig.~\ref{fig:r_and_L_CO}b), even in the most compact systems. \ion{C}{2}, \ion{Ca}{2}, and additional \ion{O}{2} lines are visible at large radii. 

Carbon and oxygen have several possible bound-bound transitions at optical wavelengths, so it is difficult to sustain high ionization in C/O-rich material and suppress optical features. Thus, C/O-rich transients are less likely to retain a featureless spectrum over a range of physical conditions, which may disfavor them as progenitors of LFBOTs and featureless TDEs. 

We note that the low-luminosity C/O-rich models ($L \lesssim 10^{42}\,\rm erg~s^{-1}$) resemble Type Ic SNe observations \citep[e.g.,][]{2023ApJ...955...71R, 2021A&A...651A..81B}. In addition, the atmospheres with higher gas temperatures — either $L \geq 10^{44}\,\rm erg~s^{-1}$ or $r_t \leq 10^{14}\,\rm cm$ — show a range of spectral behaviors, some of which are qualitatively similar to the early-time spectra of Type I SLSNe \citep{2018ApJ...855....2Q, 2019A&A...621A.141D, 2025MNRAS.541.2674A}. The models show a sequence of features from 4500 to 5000\,\AA\, including \ion{O}{2}; from 6000 to 7000\,\AA, there are some features that are also seen in SLSNe events \citep[e.g.,][]{2018ApJ...855....2Q, mazzali_2016}. However, the lines vary based on the exact conditions in the model, and the observed features also differ between events and over time. A focused analysis is necessary to study the application of these models to SLSNe. 

\subsection{Featurelessness in the Ultraviolet}

\begin{figure*}
    \centering
    \includegraphics[width=\linewidth]{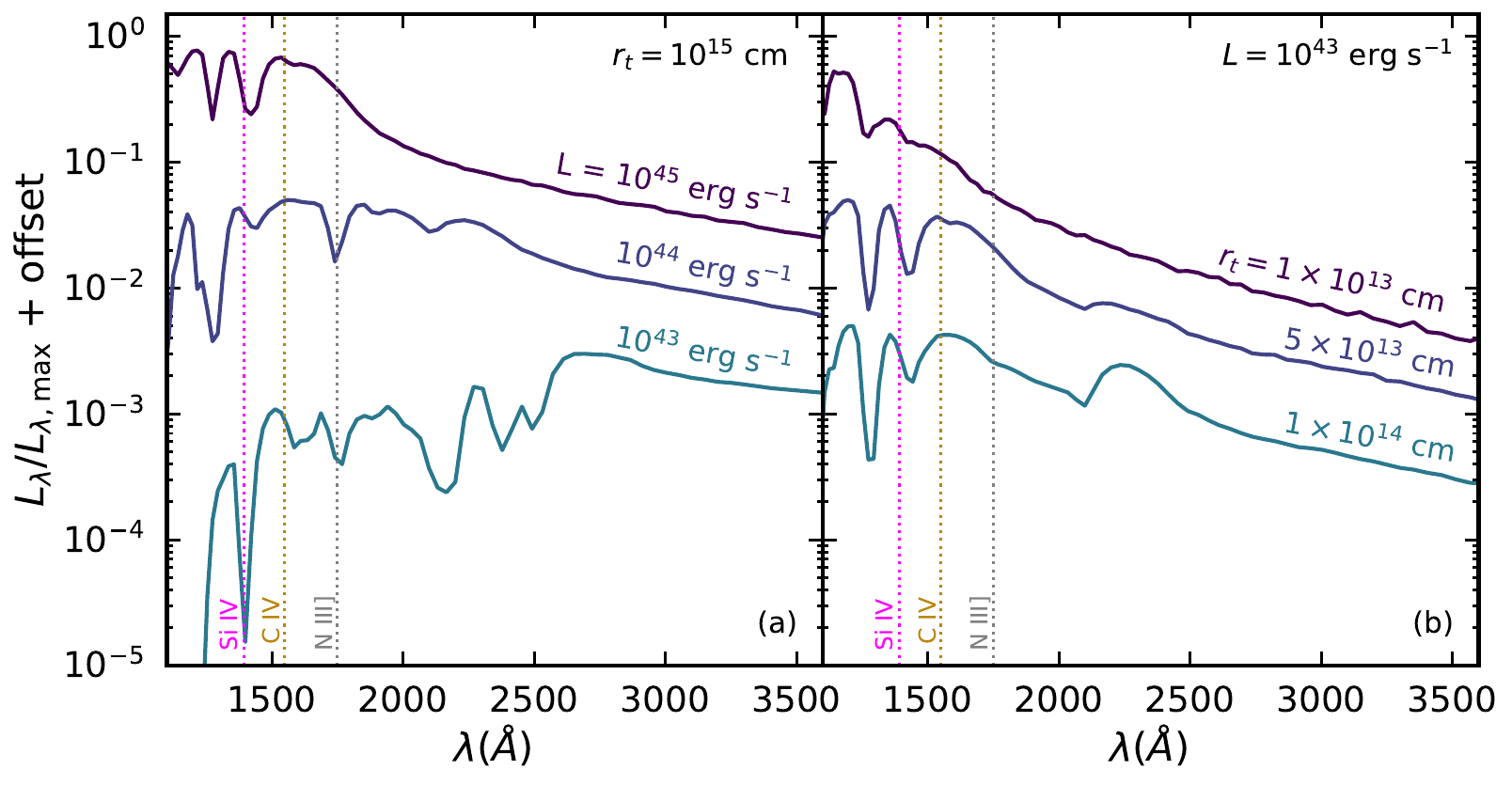}
    \caption{(a) Synthetic UV spectra from ejecta surrounding sources of varying luminosities. We use $M = 1\,M_{\odot}$ and an expansion velocity of $0.1c$. Even at the highest source luminosity, some absorption features form at UV wavelengths. (b) Synthetic UV spectra from clouds with varying characteristic radii. For each model, a $1\,M_{\odot}$ ejecta surrounds a $10^{43}$\,erg~s$^{-1}$ source and expands at $v = 0.1c$. The most compact system can emit a mostly featureless spectrum into the UV. The gas temperature remains sufficiently high up to the surface of the ejecta, where the UV continuum thermalizes.}
    \label{fig:uv_series}
\end{figure*}

Figure~\ref{fig:uv_series}a shows three spectra from 1200 to $3500\,\text{\AA}$ for a cloud expanding at $0.1c$, with $1\,M_{\odot}$ and $r_t = 10^{15}\,\rm cm$ with central sources emitting luminosities of $10^{43}$ to $10^{45}\,\rm erg~s^{-1}$. The brightest models are featureless in the near UV ($\lambda > 2000\,$\AA) but often show lines in the far UV as a result of the enhanced metal opacity at these wavelengths. Extreme radiation fields and compact radii are required to completely suppress features at wavelengths shorter than 2000\,\AA. 

Figure~\ref{fig:uv_series}b shows spectra from $1\,M_{\odot}$ ejecta with $r_t = 10^{13}\,\rm cm$, $5 \times 10^{13}\,\rm cm$, and $10^{14}\,\rm cm$ expanding at $0.1c$ around a source emitting $10^{43}\,\rm erg~s^{-1}$. Line emission weakens with decreasing cloud radius, with the spectrum reaching a near-featureless blackbody at $5 \times 10^{13}\, \rm cm$. Compact ejecta can sustain high gas temperatures near the surface of the cloud, where the UV continuum thermalizes and metal transitions may form line features. 

\section{Discussion}\label{sec:discussion}

\begin{figure}
    \centering
    \includegraphics[width=\linewidth]{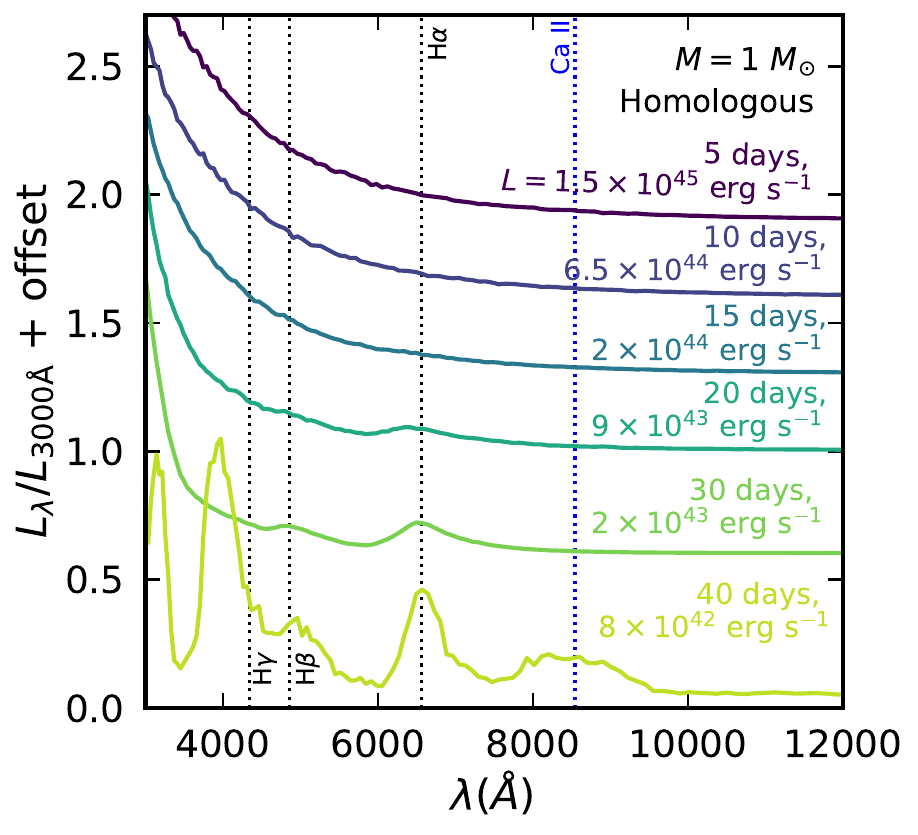}
    \caption{Synthetic spectra emitted from a $1\, M_{\odot}$ ejecta undergoing homologous expansion with AT\,2024wpp-like conditions. The luminosity evolves as $L \propto t^{-3.4}$, the medium has solar metallicity and the velocity at the power-law break is $0.1c$. By forty days, there are strong, broad emission features unlike those observed in AT\,2024wpp.
    }
    \label{fig:homologous_wpp}
\end{figure}

\begin{figure*}
    \centering
    \includegraphics[width=\linewidth]{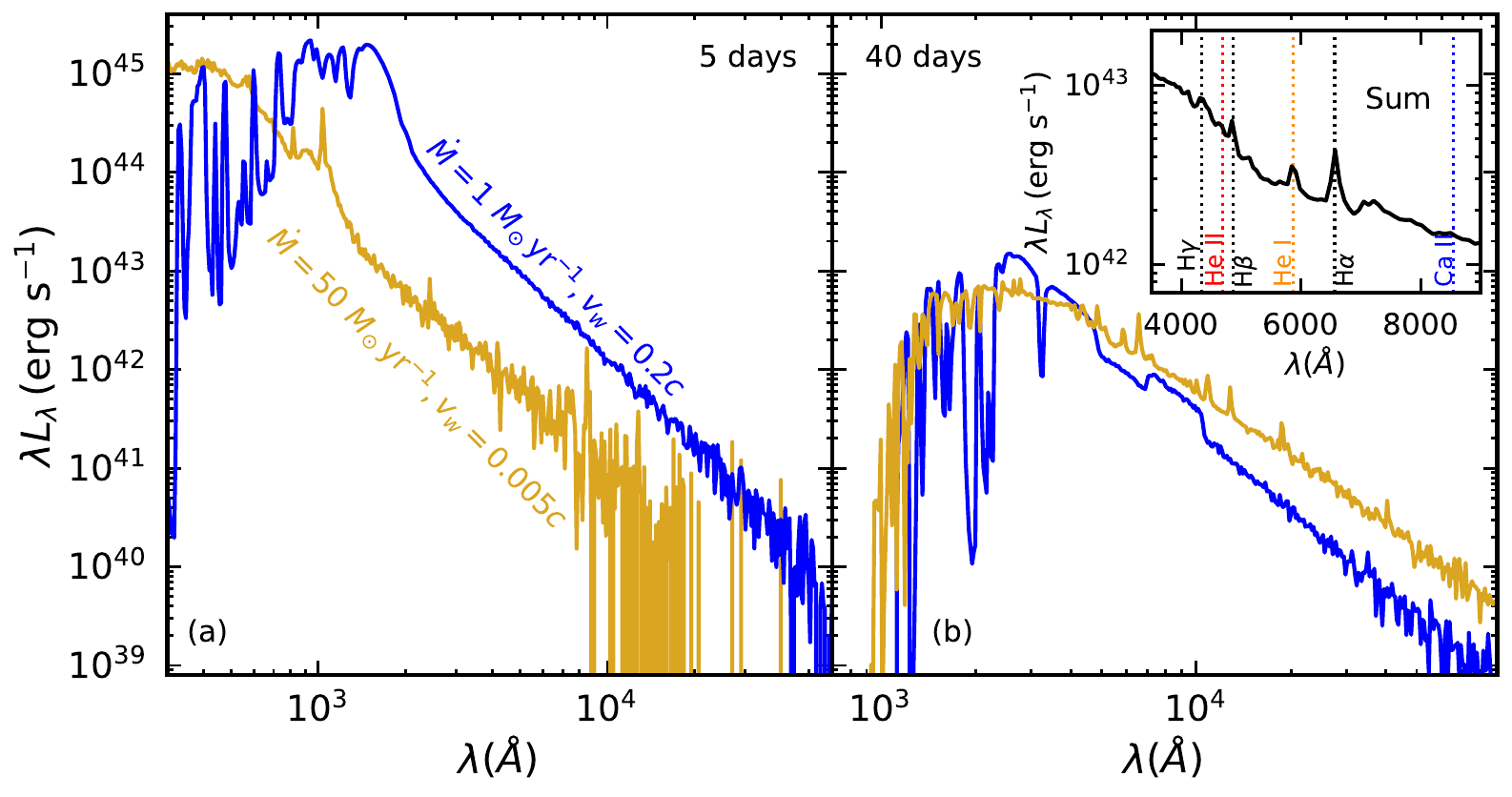}
    \caption{Synthetic spectra emitted from two outflows with constant velocities and mass-loss rates at (a) five days, with a source emitting $1.5 \times 10^{45}$~erg~s$^{-1}$ and (b) forty days, with $L = 8 \times 10^{42}$~erg~s$^{-1}$. One outflow has a mass-loss rate of $1\,M_{\odot}\,\rm yr^{-1}$ and moves at $0.2c$ (blue), while the other has $\dot{M}  = 50\,M_{\odot}\,\rm yr^{-1}$ and expands at $0.005c$ (gold). The inset shows the sum of the two forty-day spectra (black) in the optical band. At early times, the fast outflow produces bright emission, while at forty days, the spectrum from the slow-moving ejecta dominates and shows weak lines.}
    \label{fig:constant_wpp}
\end{figure*}

\subsection{Applications to AT 2024wpp and LFBOTs}


Having outlined the conditions that form or suppress spectral features in luminous transients, we consider the implications for LFBOTs such as AT\,2024wpp. During their weeks-long featureless phases, AT\,2024wpp, AT\,2018cow and CSS\,161010 sustain bolometric luminosities above $10^{43}~\rm erg~s^{-1}$ and blackbody radii between $10^{14}$ and $10^{15}$ cm \citep{margutti_cow, gutierrez_2024, 2025arXiv250900951L}, placing them in regions of suppressed line emission (Fig.~\ref{fig:space_with_contours}). We now examine what physical systems can maintain these conditions and prevent features from emerging over the observed timescales.

In some proposed progenitor scenarios for LFBOTs, a debris cloud is ejected in an impulsive, supernova-like explosion and undergoes homologous expansion \citep[e.g.,][]{2018ApJ...865L...3P, margutti_cow}. The optical and UV emission is produced when the luminosity of a central source (e.g., a magnetar or black hole) is reprocessed within this ejecta. Figure~\ref{fig:homologous_wpp} shows synthetic spectra for a homologous model ($v \propto r$) with properties based on AT\,2024wpp \citep{2025arXiv250900951L}; we use a power-law density profile with $M = 1\,M_{\odot}$ and break velocity $v_t = 0.1c$. The bolometric luminosity evolves as $L = 10^{45}~ (t/5~{\rm days})^{-\alpha}~{\rm erg~s^{-1}}$, with $\alpha =3.4$. 

At early times ($t \lesssim 20$~days) when the luminosity is high, the spectra in Figure~\ref{fig:homologous_wpp} are completely featureless — a result of both high ionization and Doppler broadening. By thirty days, however, the source luminosity has diminished, the homologous ejecta has expanded beyond $r_t = 10^{15}\,\rm cm$, and broad Balmer features begin to appear in the spectrum. After forty days, the cloud has become so extended that the photosphere begins to recede into the bulk of the ejecta; numerous prominent line features form in the cool overlying gas, and the spectrum deviates entirely from a smooth blackbody. 

A higher ejecta mass could prolong feature suppression, as the photosphere would remain in the outer layers — where the density cuts off — for a longer period before receding into the bulk. However, as the radius of this cloud increases and the luminosity drops, we expect to eventually reach the conditions for noticeable broad line emission. Moreover, a higher mass corresponds to a long diffusion time, in tension with the rapid decline of the AT\,2024wpp light curve. We conclude that a single spherically symmetric, homologous ejecta component alone is unlikely to produce the long-lived featureless blackbody spectra seen in LFBOTs. 

Similar problems face a model in which the optical and UV spectrum arises from continuous outflows launched by super-Eddington accretion onto a black hole \citep[e.g.,][]{2009MNRAS.400.2070S,2016MNRAS.461..948M}. In this picture, thermal radiation is advected with the wind and ultimately decouples at the trapping radius \citep{2009MNRAS.400.2070S}. As with a homologous flow, once the fast wind expands to a large radius, the most extended, coolest layers are likely to eventually produce conspicuous features. 
Assuming the mass loss rate of the wind, $\dot{M}$, declines over time, the optical thermalization depth will eventually recede, and the reprocessing of UV emission to optical wavelengths will become inefficient (e.g., Fig.~\ref{fig:mass_series}). 

As a solution, we could modify the spherically symmetric picture to consider an aspherical accretion disk wind model \citep[e.g.,][]{margutti_cow}, which has both a fast polar wind and a slower equatorial outflow. At early times, the optical light is dominated by the rapidly expanding polar ejecta, which appears featureless due to the high luminosities and large wind velocities. At later times, this fast ejecta becomes optically thin and the spectrum becomes dominated by reprocessed emission from the slower, compact equatorial outflow. 

Figure~\ref{fig:constant_wpp} shows synthetic spectra from a parametrized model consisting of two separate wind components with different velocities ($v_w$). This pair of one-dimensional models is meant to approximate the more complicated geometry of a multi-component wind. We set the bolometric luminosity to that inferred from observations of AT\,2024wpp at five and forty days \citep{2025arXiv250900951L}. At early times ($\sim 5$~days, Fig.~\ref{fig:constant_wpp}a), the emission from a high-velocity outflow dominates and the spectrum remains featureless as a result of the high source luminosity and strong Doppler broadening. The low velocity ejecta is so compact that it produces primarily UV and minimal optical emission. 

At later times (Fig.~\ref{fig:constant_wpp}b), the fast ejecta becomes diffuse and cool, and shows broad lines. The emission from the slow wind begins to dominate in the optical band; this ejecta component remains hot, and therefore produces a blackbody continuum with only weak Balmer and \ion{He}{1} $\lambda$5876 lines, along with very faint \ion{He}{2} $\lambda$4686. Because the expansion velocities are very low, the line profiles are broadened by electron scattering. While this highly simplified model setup is only illustrative, it suggests that an aspherical model with  fast and slow components may be able to explain the persistent blackbody and weak features observed in LFBOTs. Additional modeling in two or three dimensions is necessary to completely describe the physical picture. 

The above considerations emphasize that a key requirement for long-lived featureless spectra is to retain a compact emitting region within which the gas temperature and ionization remain high. Besides slowly expanding winds, this structure could result from an extended accretion flow or disk, or a quasi-static envelope surrounding a compact object. These radiation-pressure-supported envelopes have been discussed in the context of TDEs \citep[e.g.,][]{1997ApJ...489..573L, 2022ApJ...937L..12M}. Such atmospheres may initially expand as fallback mass is accumulated, then contract as the envelopes cool \citep{2022ApJ...937L..12M}. 

\subsection{Featurelessness in TDEs}

A population of featureless TDEs has recently been discovered \citep{hammerstein_2023, 2023ApJ...955L...6Y, 2025ApJ...993..198Y}; some jetted TDEs also show a persistent lack of optical line emission \citep{2012ApJ...753...77C, 2022Natur.612..430A}. 
Many such events \citep[e.g.,][]{hammerstein_2023, 2023ApJ...955L...6Y, 2025ApJ...993..198Y} have peak luminosities exceeding 10$^{44}~\rm erg~s^{-1}$, which is consistent with our parameter space of line suppression (Fig.~\ref{fig:space_with_contours}), and suggests that overionization is responsible in part for the absence of emission features in some TDEs. However, a subset are less luminous ($L \sim 10^{43}~\rm erg~s^{-1}$) and have smaller blackbody radii \citep[$r_{\rm bb} \lesssim 5 \times 10^{14}$ cm; e.g.,][]{2022ApJ...930...12H, 2022ApJ...937....8Y, 2025arXiv251022211Z}. Suppressing lines in these systems likely requires compact emission regions and/or high velocities.

In super-Eddington systems, the bound material from a disrupted star could inflate into a quasi-static spherical envelope that reprocesses emission from the central engine \citep{1997ApJ...489..573L}. Instead of producing extended winds, the excess accretion energy may escape through jets, allowing the atmosphere to remain bound to the central black hole \citep{2014ApJ...781...82C}. If the atmosphere is optically thick and compact, the emitted spectrum could be featureless. 

Outflows from an accretion disk could account for featureless TDE observations. In TDEs, such disk winds reprocess emission to optical wavelengths and create line features \citep{2018ApJ...859L..20D, 2022ApJ...937L..28T, 2022MNRAS.510.5426P, 2025MNRAS.540.3069P}. The bolometric luminosity from a TDE accretion disk has been shown to be a function of viewing angle, with photons preferentially escaping near the poles where the optical depth is low \citep{2025MNRAS.540.3069P}. The large luminosity emitted along the polar region of the disk can increase the ionization state of the medium and suppress lines in the outgoing spectrum. 

In the UV bands, the spectral properties of TDEs vary considerably. We do not attempt a detailed UV spectral analysis here, but note that some of our models show qualitative similarities to specific events. iPTF15af has apparent \ion{Si}{4} and \ion{C}{4} absorption features \citep{2019ApJ...873...92B}, which resemble blueshifted absorptions in our models with $L = 10^{45}\rm~erg~s^{-1}$ and $r_t = 10^{13}$ and $5 \times 10^{13}\rm~cm$. ASASSN-14li \citep{2016ApJ...818L..32C} is largely featureless to 1900\,\AA, just as our model with $L = 10^{45}\rm~erg~s^{-1}$ (Fig.~\ref{fig:uv_series}a); iPTF16fnl \citep{2018MNRAS.473.1130B} retains a smooth continuum even farther into the UV, similar to the spectrum from our most compact system (Fig.~\ref{fig:uv_series}b).  In general — given the large number of strong UV features relative to  optical lines — more extreme conditions are required to produce featureless UV spectra. In those systems where the luminosity and compactness are only modest, high velocities may be needed to wash out the line features. 

\subsubsection{Line Profiles}

Our spherically symmetric models exhibit a consistent blueshift of the emission peak that is proportional to the outward expansion velocity of the medium (Fig.~\ref{fig:velocity_profiles}a). This effect produces features with a similar shift and shape to the H$\alpha$ emission line observed in CSS\,161010 \citep{coppejans_2020, gutierrez_2024}. Other LFBOTS, such as AT2018cow, show redshifted emission peaks. These do not occur in our spherical models, and thus can be taken as an indication of an asymmetric line-forming region. At late times, AT\,2024wpp shows a double-peaked H$\alpha$ line profile that also indicates a deviation from spherical symmetry. While our low velocity wind model produces similar line strengths at these epochs (Fig.~\ref{fig:constant_wpp}), a multi-dimensional geometry is needed to explain the line profile as indicated by \citep{2025arXiv250900951L}. 

\citet{2022MNRAS.510.5426P} performed radiative transfer calculations of optical spectra from TDE accretion disks with two-dimensional, biconical winds. They find that denser outflows produce stronger features, in general agreement with our analytic results (Eq.~\ref{eqn:simple_ratio}). They further show that their outflow model can qualitatively reproduce the hydrogen and helium lines in TDE-Bowen observations.  The assumed biconical geometry does lead to double-peaked line profiles from certain viewing angles. 

\subsection{Model Limitations}

Our calculations model the line emission near the edge of an optically thick region where a quasi-blackbody continuum forms. Such an environment will reach near-LTE conditions and the line features will reflect the temperatures near the photosphere. To the extent that the radiation field deviates from blackbody, the level populations and hence line strengths will differ from what we find here; however, the qualitative relationships we have described between the strength of emission lines and the luminosity and radius of the system are likely to hold. 

If low density, extended gas lies above the photosphere, additional line formation may occur outside the region well modeled by LTE. In this case, the environment more closely resembles that of optically thin photoionized clouds, where the ionization and excitation conditions are regulated by the flux of ionizing photons rather than the local temperature. Some TDEs show spectroscopic features likely to be formed from the combination of an optically thick component and a diffuse region above the photosphere. These environments may also be relevant for LFBOTs if fast outflows grow optically thin at later times  \citep{margutti_cow, 2019ApJ...871...73H}. However, whenever a thermal continuum is observed, the conditions of line formation described above are likely to apply at least in the photospheric region. 

\citet{roth_2016} studied line formation in TDEs using a similar spherically symmetric framework. Our approach differs in a few ways. Their calculations included NLTE effects in an atmosphere composed of hydrogen, helium, carbon and oxygen, whereas we have included metals heavier than oxygen but adopted LTE for all species. \citet{roth_2016} also used a quasi-static envelope with turbulent, Gaussian line broadening, whereas we have assumed radial expansion.  

Despite these differences, the predicted line profiles are broadly similar in both cases, aside from the systematic blueshift of line peaks that arises in coherent expansion. As in \citet{roth_2016}, we find that hydrogen-rich envelopes that are sufficiently luminous and compact can produce spectra with weak hydrogen but noticeable helium lines. Although substantial H$\alpha$ emission may be generated above the thermalization depth, these photons are largely reabsorbed within the optically thick line itself or washed out by repeated electron scattering. The observed line photons are then primarily produced under the conditions near the photosphere, which our simplified treatment intends to capture. 

The inclusion of metal species heavier than oxygen can dramatically increase the opacity at UV wavelengths, primarily due to blanketing by blended iron group lines. This influences the shape of the UV continuum and the degree of reprocessing to the optical. In the calculations of \citet{roth_2016}, the UV continuum largely reflected the assumed blackbody spectrum emitted from an inner boundary condition at fixed radius, whereas we have generally placed the inner boundary below the thermalization depth at all wavelengths.    

\citet{roth_2016} found that emission lines may not be completely suppressed even at luminosities of $10^{45}\,\rm erg~s^{-1}$. While our models are generally featureless at these luminosities, the scaling relations in \S\ref{sec:analytics} show that the line strength is inversely proportional to velocity, so we expect more conspicuous features to appear in lower velocity systems. NLTE effects may also contribute to quantitative differences, and the inclusion of iron lines substantially modifies the UV spectrum, which will indirectly affect the production of optical emission lines.  

\section{Summary and Conclusions}\label{sec:conclusions}

We have studied line formation and suppression in the spectra of luminous transients. We defined the range of photospheric conditions under which H, \ion{He}{1}, and \ion{He}{2} emission lines are expected to appear, and quantified where different spectral types reside in the parameter space of luminosity and system radius. We used numerical spectral calculations to survey how source and gas properties influence line features, producing a sequence of model spectra ranging from strongly featured to nearly featureless. High source luminosities and compact ejecta radii suppress line emission, due to the high temperatures and ionization states of the emitting medium. Large velocities ($v \gtrsim 0.1c$) broaden features such that they may blend completely into the continuum. 

Our scaling relations and models predict correlations between emission line strengths and the photospheric radius and temperature. These are in general agreement with trends observed in the \ion{He}{2}/\ion{He}{1} and \ion{He}{2}/H$\alpha$ lines ratios of TDE spectra \citep{2022A&A...659A..34C}. Deviations from these trends could result from non-spherical configurations, variations in velocities, or additional line emission from low density gas under NLTE conditions.  

In expanding spherical configurations, the peaks of emission features are systematically blueshifted, an effect that is seen in some LFBOTs and TDEs. A redshifted peak or more complicated line profile shape likely implies asphericity, infall, or a more complex velocity structure. If the expansion velocity is low, electron scattering  becomes the dominant line broadening mechanism.

Because of the high UV absorptive opacity, the UV flux is modified by the cooler gas near or above the photosphere, while the optical continuum forms deeper within. As a result, a blackbody fit to optical wavelengths can over or underestimate the UV contribution depending on the conditions.
Such effects can introduce systematic errors in blackbody-derived parameters. 
Suppressing features at UV wavelengths requires more extreme conditions than in the optical, given the stronger metal lines in that band.  

The prolonged absence of features in LFBOTs is in tension with models of a single, fast homologous mass ejection, which predict strong lines to develop as the gas expands and the luminosity declines. As an alternative, we consider a multi-component model with fast polar outflows and slow-moving or quasi-static equatorial material. The highly Doppler-broadened fast outflows naturally produce an early-time featureless optical spectrum, while the slow component maintains a compact radius which at later times produces a smooth, nearly blackbody continuum with weak H$\alpha$ and \ion{He}{2} $\lambda$4686 emission, qualitatively similar to that observed in AT\,2024wpp. A similar picture may explain the prolonged lack of emission lines in featureless and jetted TDEs. Maintaining a compact emitting region, and perhaps high velocities, is likely necessary to suppress the development of optical line features as the luminosity declines.

Our modeling has focused on line formation in the photospheric regions of luminous transients, where the conditions are expected to be close to LTE. If extended low density gas lies above this region, photoionization or interaction shocks may generate additional line emission with qualitatively different behaviors. 
Future work will incorporate multi-dimensional and NLTE treatments to more fully capture non-thermal processes and orientation-dependent effects. Spectra deviating from the photospheric trends identified here may be diagnostic of the distinct emission sites in these systems. 

\begin{acknowledgements}
We thank the anonymous referee for helpful comments that improved this manuscript. We thank Wenbin Lu, Raffaella Margutti, Yuhan Yao, Ryan Chornock and Natalie LeBaron for useful conversations. 
This material is based upon work supported by the National
Science Foundation under Grant No.~2206713. DK is supported in part by the U.S. Department of Energy, Office of Science, Office of Nuclear Physics, DE-AC02-05CH11231, DE-SC0004658, and DE-SC0024388, and by a grant from the Simons Foundation (622817DK). This work benefited from discussions at workshops supported by the Gordon and Betty Moore Foundation through grant GBMF5076. 
\end{acknowledgements}

\vspace{5mm}

\software{\texttt{Sedona} \citep{2006ApJ...651..366K}, numpy \citep{2020Natur.585..357H}, matplotlib \citep{2007CSE.....9...90H}, h5py}

\bibliography{sample631}{}
\bibliographystyle{aasjournal}



\end{document}